\documentclass{nature3}

\usepackage{amssymb}
\usepackage{amsmath}
\usepackage{bm}
\usepackage{comment}
\usepackage{journaux}
\usepackage{graphicx}

\usepackage{floatrow}

\title{Obliquity-Driven Sculpting of Exoplanetary Systems}

\author{Sarah Millholland$^{1,2}$ and Gregory Laughlin$^{1}$}

\begin{document}

\maketitle

\begin{affiliations}
\item Department of Astronomy, Yale University, New Haven, CT 06511
\item NSF Graduate Research Fellow
\end{affiliations}


\textbf{NASA's Kepler mission revealed that $\sim 30\%$ of Solar-type stars harbor planets with sizes between that of Earth and Neptune on nearly circular and co-planar orbits with periods less than 100 days\cite{2013ApJS..204...24B, 2013ApJ...766...81F, 2013PNAS..11019273P, 2018ApJ...860..101Z}. Such short-period compact systems are rarely found with planet pairs in mean-motion resonances (MMRs) -- configurations in which the planetary orbital periods exhibit a simple integer ratio -- but there is a significant overabundance of planet pairs lying just wide of the first-order resonances\cite{2011ApJS..197....8L}.
Previous work suggests that tides raised on the planets by the host star may be responsible for forcing systems into these configurations by draining orbital energy to heat\cite{2010MNRAS.405..573P, 2012ApJ...756L..11L,2013AJ....145....1B}. Such tides, however, are insufficient unless there exists a substantial and as-yet unidentified source of extra dissipation\cite{2013ApJ...774...52L,2015MNRAS.453.4089S}. Here we show that this cryptic heat source may be linked to ``obliquity tides'' generated when a large axial tilt (obliquity) is maintained by secular resonance-driven spin-orbit coupling. We present evidence that typical compact, nearly-coplanar systems frequently experience this mechanism, and we highlight additional features in the planetary orbital period and radius distributions that may be its signatures. Extrasolar planets that maintain large obliquities will exhibit infrared light curve features that are detectable with forthcoming space missions. The observed period ratio distribution can be explained if typical tidal quality factors for super-Earths and sub-Neptunes are similar to those of Uranus and Neptune.}

The statistical excess of planets wide of first-order MMRs must be linked to the nature and extent of the mass accretion, orbital migration, and energy dissipation that characterized the planetary formation phases; existing explanations of the excess are connected to one or more of these processes\cite{2013ApJ...770...24P,2014ApJ...795...85W, 2017A&A...602A.101R, 2017AJ....154..236W}. Tidal energy dissipation drives planetary spins toward zero obliquity and an equilibrium rotation period close to the orbital period\cite{1981A&A....99..126H}. Maintenance of a non-zero obliquity in the face of dissipation requires an external driving force, which may be attained if the planet is locked in a secular spin-orbit resonance comprising synchronous precession of the planetary spin and orbital angular momentum vectors\cite{2007ApJ...665..754F}. Such a resonance is an instance of a ``Cassini state'', an equilibrium configuration of the spin vector\cite{1966AJ.....71..891C, 2015A&A...582A..69C}. There are four Cassini states, and Cassini state 2 is most favorable for maintaining a large obliquity in the presence of tides. When locked in a high-obliquity state, energy dissipation in the planet is tremendously enhanced, for example by a factor of $\sim$1,000-10,000 according to traditional equilibrium tide theory in the viscous approximation\cite{2007A&A...462L...5L, 2008Icar..193..637W}. A dissipative Cassini state can be difficult to maintain for hot Jupiters\cite{2007A&A...462L...5L,2007ApJ...665..754F}, but we will show it is significantly easier for planets in the systems under consideration here. 

Establishment of resonance can occur if a planet's spin precession period, $T_{\alpha} = 2\pi/(\alpha\cos\epsilon)$, where $\alpha$ is the spin-axis precession constant and $\epsilon$ the obliquity, and the period of its orbit nodal recession, $T_g = 2\pi/\lvert g \rvert$, where $g = \dot{\Omega}$, evolve to equality starting with $T_{\alpha}/T_g > 1$. 
In a multiple-planet system, nodal recession arises as a consequence of secular perturbations between pairs of planets on mutually inclined orbits (Figure 1). 
Since $g$ and $\alpha$ depend on the semi-major axes, a natural mechanism for resonant capture and obliquity excitation stems from evolution of $T_{\alpha}/T_g$ driven by protoplanetary disk-driven migration\cite{2010exop.book..347L}. Under such conditions, capture into secular spin-orbit resonance can occur if the frequency crossing condition is met, provided that the migration timescale significantly exceeds the spin-axis precession period\cite{2004AJ....128.2501W}. 
The opportunities for frequency crossings are more abundant in systems with more than two planets, since there are a greater number of orbital precession frequencies available for spin-orbit commensurability. Resonance can readily be established even though the orbital precession is a non-uniform superposition of several modes. For example, Saturn's $27^{\circ}$ axial tilt is attributed to resonant interaction with the precession of Neptune's orbital node, despite the fact that Neptune's influence on Saturn is more than $20\times$ weaker than that provided by Jupiter and Uranus\cite{2004AJ....128.2501W}.

\begin{figure}[H]
\includegraphics[width=0.93\textwidth]{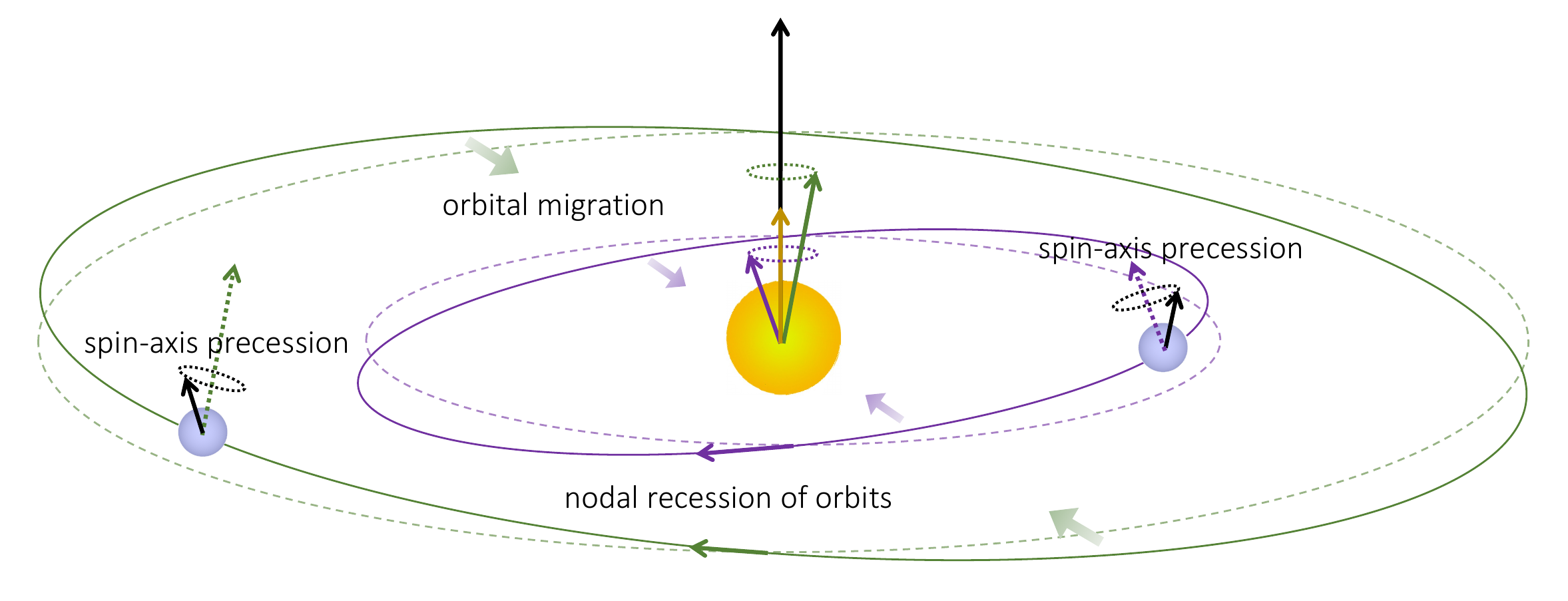}
\caption{\textbf{Schematic representation of the excitation of planetary obliquities through secular spin-orbit resonant interaction.} \newline
Two planets with a modest mutual inclination (exaggerated in this diagram) undergo mutual nodal recession of their orbits due to secular perturbations. We assume for simplicity that the stellar spin angular momentum vector is aligned with the planets' total orbital angular momentum vector. The planets' spin axes precess about their respective orbit normal vectors due to torques from the star on the planets' oblate figures. As orbital migration ensues, the nodal recession and spin-axis precession frequencies evolve and can become commensurable, at which point large obliquities can be excited through capture and subsequent evolution in spin-orbit resonance.}
\label{Geometric set-up}
\end{figure}

If systems frequently have planets locked in Cassini state 2, then $\lvert g \rvert \sim \alpha$ should be common; here we show this is the case. We consider a set of 145 planets in 55 \textit{Kepler} multiple-planet systems from Hadden \& Lithwick (2017)\cite{2017AJ....154....5H}. Using a minimal number of assumptions (see Methods), we calculated the spin-axis precession constants of the planets in the sample. The results are displayed in the left panel of Figure 2. Also included in this panel is a histogram of the Laplace-Lagrange nodal recession frequencies, $\lvert g \rvert$, for all sets of within-system pairs of planets in the 55 systems. The distributions of $\lvert g \rvert$ and $\alpha$ have a remarkably similar center and range, indicating that the two frequencies are intrinsically commensurable. The favorable frequency commensurability is far more explicit when the proximity to spin-orbit resonance is calculated for each planet individually. We detail this procedure in the Methods and present the results in the right panel of Figure 2.

\begin{figure}[H]
\includegraphics[width=0.95\columnwidth]{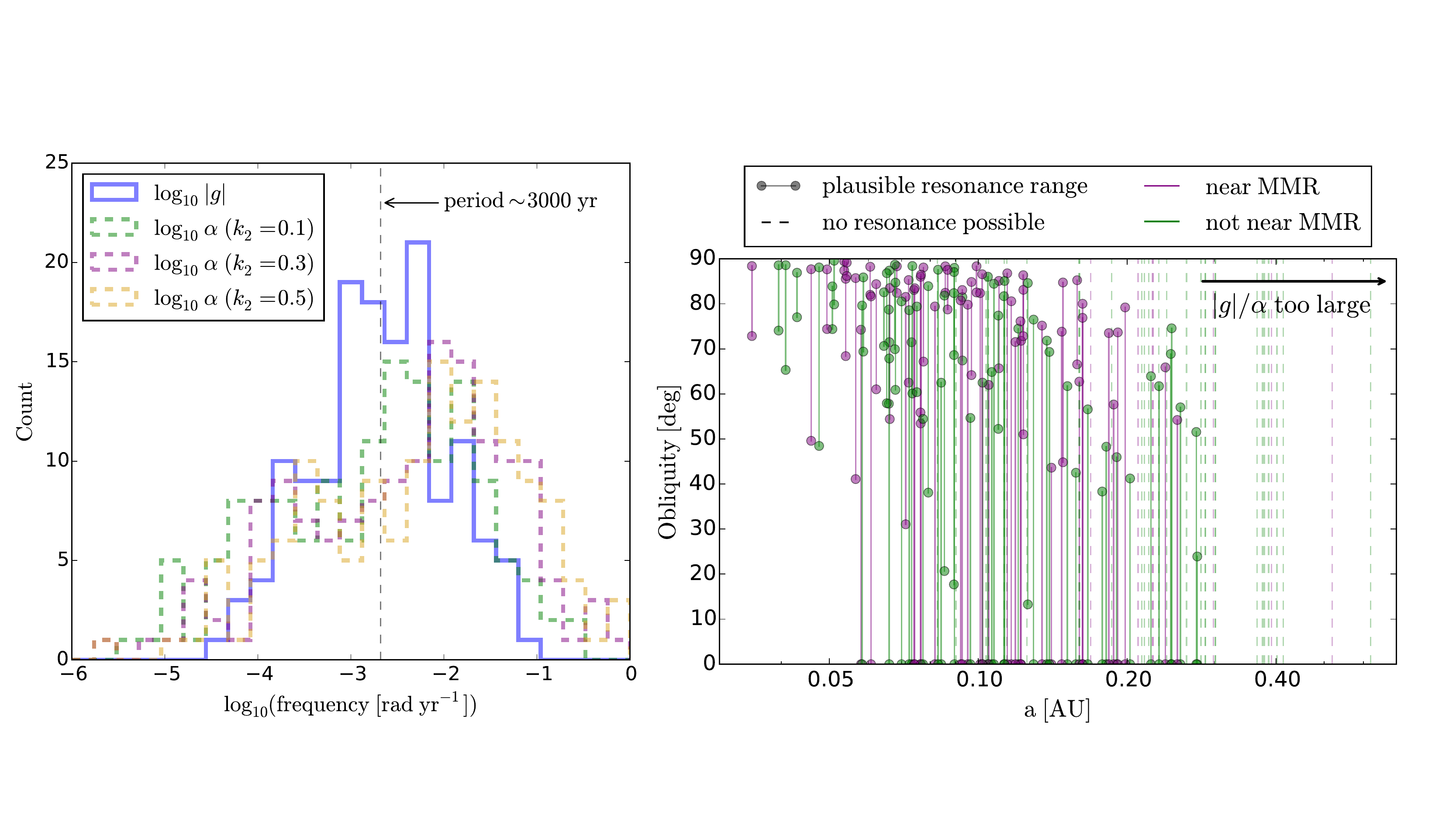}
\caption{\textbf{Intrinsic frequency commensurability for typical compact extrasolar systems.} \newline 
We illustrate the proximity to spin-orbit resonance for a sample of 145 planets in 55 \textit{Kepler} multiple-planet systems. \textit{Left panel}: The blue histogram shows the Laplace-Lagrange nodal recession frequencies, $\lvert g \rvert = \dot{\Omega}$, for all sets of within-system pairs of planets in the 55 systems. The green, purple, and yellow histograms show estimates of the the spin-axis precession constants, $\alpha$, for all planets in the sample. \textit{Right panel}: Corresponding to each of the 145 planets in the sample, the lines depict the range of plausible obliquities the planets could have if they are in spin-orbit resonance. Dotted lines are used when the resonance is unlikely for that planet. $\sim80\%$ of planets and $\sim90\%$ of those in near-MMR systems could plausibly be in resonance.}
\label{Kepler multis}
\end{figure}

Because $\lvert g \rvert \sim \alpha$ appears common among known exoplanets, secular spin-orbit resonant encounters and their associated obliquity excitations should frequently occur whenever $g$ and $\alpha$ evolve. In particular, for planets in systems wide of first-order MMRs, $T_{\alpha}/T_g$ must have changed substantially as convergent orbital migration drove the systems toward the observed orbital period commensurability. We illustrate this scenario by constructing direct and secular models of the orbital and spin evolution of a young planetary system that migrates into simultaneous spin-orbit and mean-motion resonances. Model details are provided in the Methods, but the basic set-up consists of a star and two planets, with all three bodies endowed with structure. We assume the planets are initially wide of a $(k+1):k$ orbital period ratio and that they migrate inwards convergently due to interactions with the protostellar disk. We include accelerations on the planets due to the star's quadrupole gravitational potential, and account for tides raised on the planets by the host star using equilibrium tide theory in the viscous approximation\cite{1981A&A....99..126H}. Of course, the details of the simulation depend on this choice of tidal model. Alternative models are discussed in the Methods.

Figure 3 displays dynamical evolution of the type we propose. Two prototypical sub-Neptune planets undergo convergent migration with timescales, $\tau_{a_1}=a_1/\dot{a}_1=5\,{\rm Myr}$ and $\tau_{a_2}=\tau_{a_1}/1.1$. The initial orbits are placed wide of the 3:2 MMR; the planets are initiated with negligible obliquities, $\epsilon_1$ and $\epsilon_2$, and spin periods reflecting the partial initial spin-orbit synchronization expected from tidal evolution acting on rapidly spinning birth states. While we have adopted a specific parameter configuration, we emphasize that no fine tuning was carried out. The key features of the subsequent evolution are robust.

Initial development of the system is characterized by the simultaneous decrease of $T_g$ arising from the decaying orbits and the evolution of $T_{\alpha_i}$ from migration and ongoing spin decay. 
After $\sim 8\times10^{5} \ \mathrm{yr}$, $T_{\alpha_1}/T_g$ crosses unity from below, producing a near-impulsive resonant kick that raises $\epsilon_1$ to $\sim30^{\circ}$. 
Strongly enhanced dissipation within the planet then begins to gradually right its spin axis. Once the differentially decaying orbits are captured into the 3:2 MMR, the ensuing resonant interaction increases $T_g$ and precipitates a crossing through $T_{\alpha_1}/T_g =1$ from above, eliciting capture into Cassini state 2.
The inner planet's obliquity is forced to $\epsilon_1=52^{\circ}$ by the time when the imposed migration ends. At this point, the system is set to evolve in a long-term, quasi-steady state toward a final period ratio, $P_2/P_1>1.5$.
Obliquity-enhanced tidal damping generates a luminosity, $\dot{E_1} \approx 3 \times 10^{22}\,{\rm erg} \ {\rm s}^{-1}$, enabling evolution to $\Delta=2P_2/3P_1 - 1 = 0.05$ (a resonant offset typical to the observed systems) within $\sim 7\,{\rm Gyr}$. As the period ratio grows, the planets' mutual inclination must decrease such that the system conserves total angular momentum. The requisite decrease is not extreme and can be accounted for by modest, few-degree initial mutual inclinations.

The previous simulation demonstrated the efficacy of migration-driven spin-orbit resonant capture for a prototypical short-period compact system. This type of simulation is expensive, however, and impractical for obtaining a complete portrait of the mechanism's operable domain in parameter space. We may construct such a mapping of parameter space by reducing the space to two principal parameters. Towards this goal, we consider a pair of planets that start with $P_2/P_1$ wide of first-order MMR and migrate inwards convergently for $t_f$ years. Inward migration generally decreases the ratio $T_{\alpha}/T_g = \lvert g \rvert/(\alpha\cos\epsilon)$, and resonant crossing requires that the ratio pass through unity from above. Whether crossing occurs therefore depends largely on the planets' semi-major axis evolution, so we take $a_1(t=t_f)$ as the first of the two principal parameters. Secondly, resonant capture requires that the crossing is adiabatic, in other words, that the passage through resonance is slow in comparison to the resonant libration period\cite{2004AJ....128.2510H}. Accordingly, we take the migration timescale, $\tau_{a_1}=a_1/\dot{a}_1$, as the second of the two principal parameters.

\begin{figure}[H]
\includegraphics[width=0.8\columnwidth]{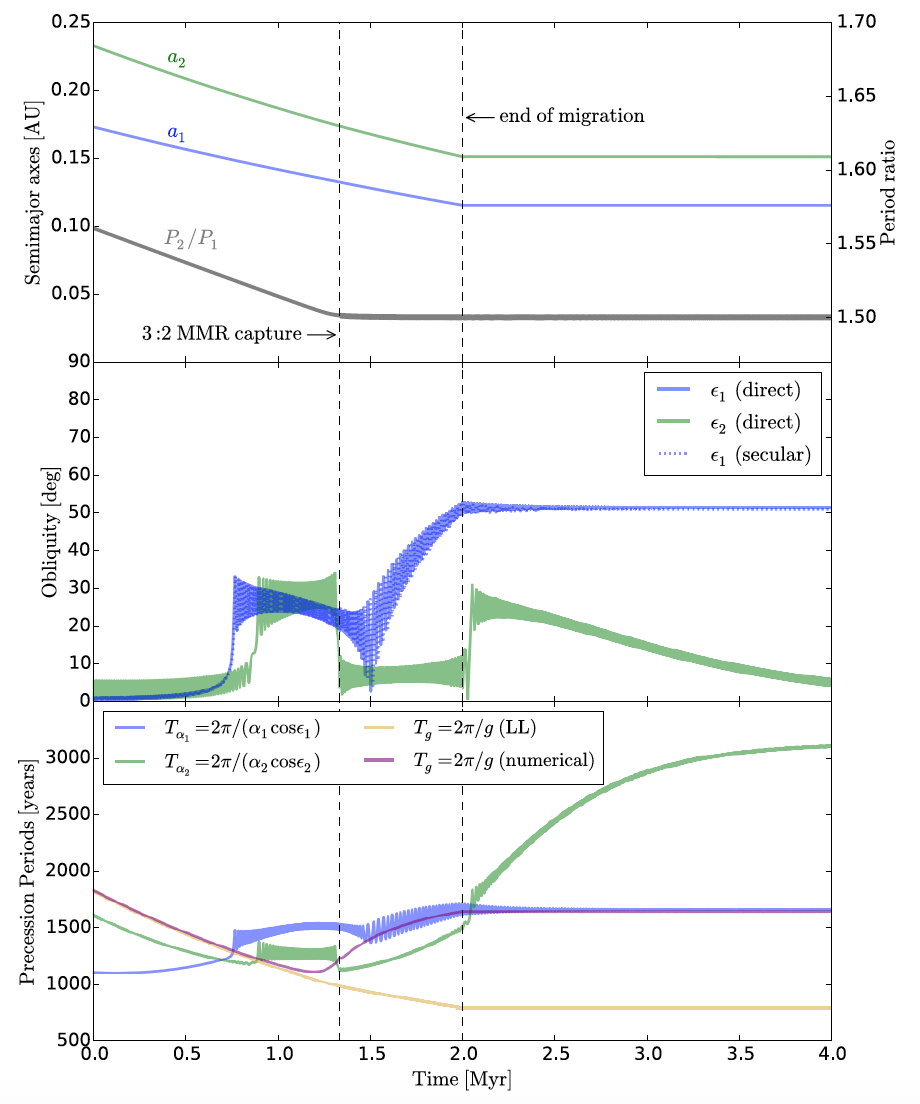}
\caption{\textbf{Capture into simultaneous mean-motion and spin-orbit resonances.} \newline
Dynamical evolution of a fiducial, two-planet system undergoing (imposed) inward, convergent orbital migration for the first $2 \ \mathrm{Myr}$. \textit{Top panel}: Evolution of the two planet's semi-major axes and period ratio. The 3:2 MMR is captured at $\sim 1.3 \ \mathrm{Myr}$. The planets' eccentricities get excited to $\sim0.04$ during the capture and ensuing migration. \textit{Middle panel}: Evolution of the two planets' obliquities, with both the direct and secular codes. \textit{Bottom panel}: The blue and green curves represent the precession periods of the inner and outer planets, respectively. The yellow curve indicates the nodal recession period according to Laplace-Lagrange theory, and the purple curve shows the true nodal recession period, which increases as the 3:2 MMR is approached. While the outer planet undergoes several resonant kicks but no captures, the inner planet experiences a resonant kick at $\sim 0.8 \ \mathrm{Myr}$ and capture into Cassini state 2 at $\sim 1.5 \ \mathrm{Myr}$. It is subsequently forced to a $52^{\circ}$ obliquity. This highly dissipative state is maintained indefinitely. The parameters of the simulation are the following: $M_{\star} = M_{\odot}$, $R_{\star} = R_{\odot}$, $P_{\star} = 20 \ \mathrm{days}$, $k_{2,\star} = 0.1$, $C_{\star} = 0.07$, $m_1 = m_2 = 5 M_{\oplus}$, $R_1 = R_2 = 2.5 R_{\oplus}$, $k_{2,1} = k_{2,2} = 0.4$, $C_1 = C_2 = 0.25$, $Q_{1,0} = Q_{2,0} = 10^4$, $e_{1,0} = e_{2,0} = 0.01$, $(P_{\mathrm{rot}})_{1,0} = 5 \ \mathrm{days}$, $(P_{\mathrm{rot}})_{2,0} = 3 \ \mathrm{days}$, $\epsilon_{1,0} = \epsilon_{2,0} = 1^{\circ}$. }
\label{fiducial simulation}
\end{figure}

Figure 4 displays the inner planet's resonant capture domain in $a_1(t=t_f)$ and $\tau_{a_1}$ space by adopting the physical parameters of the above simulation. The details of the figure's construction are provided in the Methods. The solid contours show where $(T_{\alpha}/T_g)_{t=0} > 1$ and $(T_{\alpha}/T_g)_{t=t_f} < 1$ for select values of the obliquity at $t=0$ and the MMR-induced modification to the Laplace-Lagrange nodal regression frequency, $g/g_{\scriptscriptstyle \mathrm{LL}}$. For the 2:1 and 3:2 MMRs, this ratio is always less than one; for example, $g/g_{\scriptscriptstyle \mathrm{LL}} \approx 0.5$ in the Figure 3 simulation. We observe that the domain for resonant capture indicated by the arrows corresponds exactly to the regime where close-in compact systems are found: $a\sim0.05-0.15 \ \mathrm{AU}$. Moreover, the resonant region broadens appreciably for $g/g_{\scriptscriptstyle \mathrm{LL}} < 1$.  Planets that encounter MMR are therefore more prone to spin-orbit resonances that lock their obliquities at large values and induce accelerated tidal evolution. The details of the above resonant capture map depend on the system parameters we have adopted. For instance, the resonant region is larger for the 2:1 resonance and for larger planet radii (see the supplementary figures for additional examples). The general features of the map, however, are universal.

\begin{figure}
\includegraphics[width=0.7\textwidth]{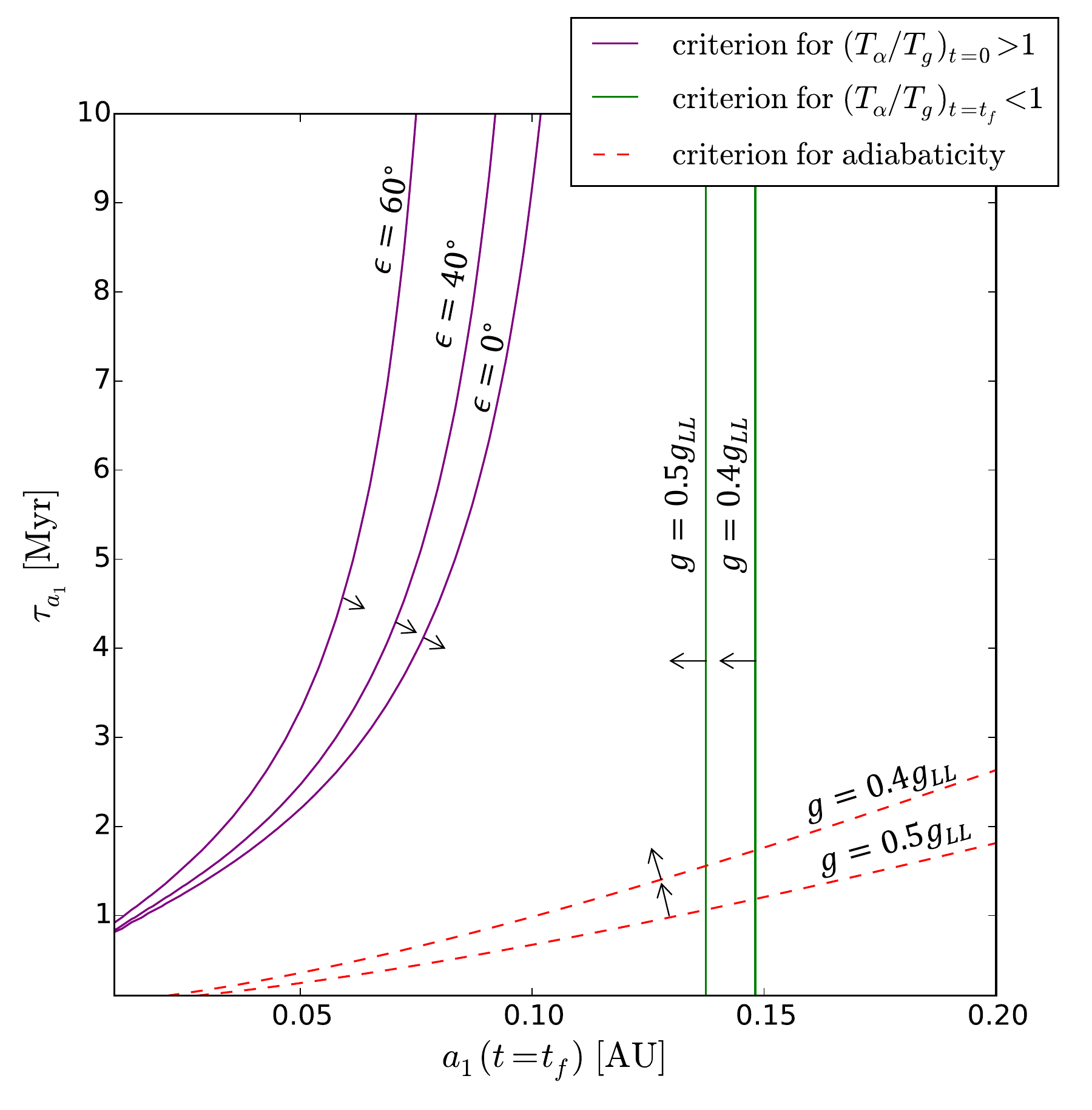}
\caption{\textbf{Domain for the inner planet's spin-orbit resonant capture during convergent inward migration.} \newline 
The approximate domain in $a_1(t=t_f)$ and $\tau_{a_1} = a_1/\dot{a_1}$ space where capture is possible for the inner planet. We use the same stellar and planetary physical parameters as the Figure 3 fiducial simulation. We assume equilibrium rotation rates, fix $(a_2/a_1)_{t=t_f} = 1.5^{2/3}$ (3:2 MMR) and $(a_2/a_1)_{t=0} = 1.05(a_2/a_1)_{t=t_f}$, and set the end of the migration time to be $t_f = 2 \ \mathrm{Myr}$. Resonant crossing requires that $T_{\alpha}/T_g$ -- which decreases during inward migration -- pass through unity from above. For a given obliquity at $t=0$, to the right of the purple line shows where $(T_{\alpha}/T_g)_{t=0} > 1$. Similarly, to the left of the green line shows where $(T_{\alpha}/T_g)_{t=t_f} < 1$ for a given MMR-induced modification to the Laplace-Lagrange secular nodal frequency. Finally, resonant capture requires that the crossing is adiabatic; the domain above the red dashed line satisfies this criterion. Note that the resonant domain is larger for the 2:1 MMR and for planets with larger radii. (See examples in the supplementary figures.)}
\label{Parameter space map}
\end{figure}

The foregoing models assumed the presence of a fully-formed parent star. In reality, however, pre-main sequence stars are rapidly rotating and have large radii. The large gravitational quadrupole moment of the rotationally flattened young star induces an additional component to each planet's orbit nodal recession. As the star spins down and contracts, transient resonant locks and obliquity excitations are readily produced, temporarily increasing $T_{\alpha}$ and substantially increasing the probability that $T_{\alpha}/T_g$ will later evolve through unity from above. The stellar evolution thus ultimately increases the likelihood that one or both planets wind up in a long-lived high-obliquity state. A version of the Figure 3 simulation that includes stellar spin-down is discussed at length in the Methods.

The interplay between MMRs, secular orbital interactions, stellar spin-down, disk-driven migration, and enhanced long-acting dissipation maintained by secular spin-orbit resonance is both complex and broadly consequential for the evolution of planetary systems. Three predictions of obliquity tide-driven resonant repulsion are particularly noteworthy and, in fact, can clearly be seen in the data. First, if tidal dissipation explains the pile-up of systems wide of first-order MMR, then the systems with the smallest orbital periods should have the largest offsets from exact resonance. This key clue has already been confirmed in the data\cite{2014A&A...570L...7D, 2017A&A...602A.101R}. Secondly, as mentioned previously, the domain for spin-orbit resonant capture is larger for the 2:1 MMR than for 3:2. This agrees with the \textit{Kepler} observation that the 2:1 MMR has a more pronounced pile-up of systems wide of resonance. Finally, the resonant capture domain is larger for planets with larger radii, and moreover, for a given tidal quality factor, $Q$, the tidal friction timescale is $t_F\propto(a/R_{\rm p})^5$. If our theory is correct, we expect larger average radii for planets in pairs with period ratios wide of MMR compared to those just inside MMR. This effect is present in the population of planets detected by \textit{Kepler} (see Methods). Among the 150 planets in pairs closest to the inside of the 3:2 and 2:1 commensurabilities, the median radius is 0.14$R_{\rm Jup}$, whereas for planets in pairs just outside, the median increases to 0.21$R_{\rm Jup}$. Mood's median statistical hypothesis test shows the difference to be significant with p-value, $p < 6\times10^{-5}$. Furthermore, planets just outside resonance also show a systematically steeper $R_{\rm p}(a)$ relation. NASA's TESS Mission\cite{2015JATIS...1a4003R} will substantially augment the statistics of this population and will permit masses for many additional planets in near-resonant pairs to be measured.

An additional implication concerns constraints on the planets' tidal quality factors, $Q$. For planets undergoing resonant repulsion, the distance traveled beyond resonance is related to the amount of dissipation experienced by the planets over the system age, which in turn is directly connected to $Q$. This relation may be used to obtain order-of-magnitude constraints on the $Q$ values of planets in compact multiple-planet systems. We present such a calculation in the Methods. The resulting distribution of $Q$ estimates peaks at $Q\sim10^4$, suggesting that planets in these compact exoplanetary systems may have typical dissipation efficiencies similar to those of Uranus and Neptune.

Secular spin-orbit coupling may have significant consequences for compact exoplanetary systems, and this points toward areas for additional investigation. Exploratory calculations (A.D. Adams et al. manuscript in preparation) indicate that oblique, transiting, short-period planets show distinctive signatures in full phase infrared light curves that will be discernible with Spitzer or JWST\cite{2016MNRAS.457..926S}. Another application pertains to planets that have undergone differential migration to form resonant chains, such as the seven-planet system orbiting TRAPPIST-1, which may be particularly susceptible to establishment of spin-orbit resonances due to the wide range of sweeping forcing frequencies during the assembly process\cite{2017ApJ...840L..19T}. For many-planet systems like TRAPPIST-1, chaotic regions in the spin dynamical phase space will likely arise due to resonance overlap (see the Methods). Any large axial tilts that result from either resonance capture or chaos will preclude synchronous rotation, and obliquity tides will generate strong interior heating, leading to significant meteorological and geophysical consequences\cite{1997Icar..129..450J, 1979Sci...203..892P}.


\subsection{\large{References}}~\\

\section*{\large{Acknowledgements}} 
We thank Konstantin Batygin, Dan Fabrycky, and Yanqin Wu for inspiring conversations. S.M. is supported by the National Science Foundation Graduate Research Fellowship Program under Grant Number DGE-1122492. This material is also based upon work supported by the National Aeronautics and Space Administration through the NASA Astrobiology Institute under Cooperative Agreement Notice NNH13ZDA017C issued through the Science Mission Directorate. We acknowledge support from the NASA Astrobiology Institute through a cooperative agreement between NASA Ames Research Center and Yale University.

\section*{\large{Author Information}}
\subsection{Affiliations} 
\textit{Department of Astronomy, Yale University, New Haven, CT, USA}
Sarah Millholland and Gregory Laughlin 
\subsection{Contributions} 
S.M. performed the simulations, calculated the resonant proximity diagnostics, and made the resonant capture parameter space map. G.L. conceived of and derived the constraints on the tidal quality factors, the radius distribution observations, and the predictions regarding satellites. Both authors wrote the paper and made figures.
\subsection{Competing interests}
The authors declare no competing financial interests.
\subsection{Corresponding author}
Correspondence and requests for materials should be addressed to Sarah Millholland (sarah.millholland@yale.edu).

\clearpage

\section*{\Large{Methods}}

The Methods section is divided into two parts for clarity. The first part contains the details of the findings and figures presented in the main text. The second part contains information that is supplemental in nature, including background material and broader implications of the theory.

\subsection{\large{Part 1: Details of main text calculations}}

\subsection{1.1 Susceptibility of compact exoplanetary systems to spin-orbit resonance.} 

In the main text, we showed that $\lvert g\rvert \sim \alpha$ is common for typical \textit{Kepler} short-period, multiple-planet systems (Figure 2). Here we provide the details of this computation.

We begin by defining the relevant resonant frequencies, $\alpha$ and $\lvert g \rvert$. The torque from the host star on a rotationally-flattened planet will cause the planet's spin-axis to precess about the orbit normal (Figure 1) at a period, $T_{\alpha} = 2\pi/(\alpha\cos\epsilon)$. Here $\epsilon$ is the obliquity and $\alpha$ is the precession constant\cite{2004AJ....128.2501W}, which, in the absence of satellites, is given by  

\begin{equation}
\alpha = \frac{1}{2}\frac{M_{\star}}{m_{\mathrm{p}}}\left(\frac{R_{\mathrm{p}}}{a}\right)^3\frac{k_2}{C}\omega.
\label{alpha}
\end{equation}
Here $M_{\star}$ is the stellar mass, $m_{\mathrm{p}}$ the planet mass, $R_{\mathrm{p}}$ the planet radius, $a$ the semi-major axis, $k_2$ the Love number, $C$ the moment of inertia normalized by $m_{\mathrm{p}} {R_{\mathrm{p}}}^2$, and $\omega$ the spin frequency. We have assumed that $J_2$, the coefficient of the quadrupole moment of the planet's gravitational field, takes the form\cite{2009ApJ...698.1778R},
\begin{equation}
J_2 = \frac{\omega^2 {R_{\mathrm{p}}}^3}{3 G {m_{\mathrm{p}}}}k_2.
\end{equation}
Order-of-magnitude estimates of $\alpha$ are fairly insensitive to $k_2$ and $C$. In the Solar System, typical values for both $k_2$ and $C$ are of order 0.2 to 0.4, whereas $\alpha$ varies dramatically, ranging from $2\pi/\alpha=2.6\times10^{4}\,{\rm yr}$ for Earth to $2\pi/\alpha=2.3\times10^{7}\,{\rm yr}$ for Neptune. A simplification results from eliminating $C$ in equation (\ref{alpha})
with the Darwin-Radau approximate relation\cite{2011A&A...528A..18K}, 
\begin{equation}
C = \frac{2}{3}\left[1-\frac{2}{5}\left(\frac{5}{k_2 + 1} - 1\right)^{1/2}\right].
\label{Darwin-Radau}
\end{equation}

In addition to the spin-axis precession constant, the other relevant frequency for secular spin-orbit resonant capture is provided by a planet's orbit nodal regression, $g = \dot{\Omega}$. Nodal regression may arise due to a variety of dynamical influences; one is secular perturbations between planets in a multiple-planet system (Figure 1). In a two-planet system, the nodes of both planets regress uniformly. The frequency is given by
\begin{equation}
g_{\scriptscriptstyle \mathrm{LL}} = -\frac{1}{4} b_{3/2}^{(1)}(\alpha_{12})\alpha_{12}\left(n_1\frac{m_2}{M_{\star} + m_1}\alpha_{12} + n_2\frac{m_1}{M_{\star} + m_2}\right)
\label{LL g}
\end{equation} 
if the planets are not near mean-motion resonance\cite{1999ssd..book.....M} (where the LL subscript stands for Laplace-Lagrange).  Here, $\alpha_{12} = a_1/a_2$ and $n_i$ is the mean-motion of planet $i$, $n_i^2 = G M_{\star}/a_i^3$. The constant, $b_{3/2}^{(1)}(\alpha_{12})$ is a Laplace coefficient, defined by 
\begin{equation}
b_{3/2}^{(1)}(\alpha_{12}) = \frac{1}{\pi}\int_{0}^{2\pi}\frac{\cos\psi}{(1-2\alpha\cos\psi + \alpha^2)^{3/2}}d\psi.
\end{equation} 
Modification of $g_{\mathrm{LL}}$ due to stellar and planetary oblateness is negligible, except during the star's pre-main sequence phase (see the Methods).  If the planets are near MMR, the frequency is smaller by a factor of $g/g_{\scriptscriptstyle \mathrm{LL}}\sim 0.5$, where $g_{\scriptscriptstyle \mathrm{LL}}$ is equation (\ref{LL g}). (In practice, the resonant modification to the Laplace-Lagrange frequency given by equation (\ref{LL g}) is determined numerically\cite{2010A&A...511A..21C}.) 

In systems with three or more planets, each planet's node will not regress uniformly, but rather will have several modes to its perturbation associated with interactions with the other $n_{\mathrm{pl}} -1$ planets in the system. The frequencies of these modes are close but not exactly equal to the pairwise applications of equation (\ref{LL g}) and rather must be calculated with full Laplace-Lagrange theory. However, the close approximation provided by equation (\ref{LL g}) is sufficient for our analyses. The amplitude of each component depends on the planet masses and semi-major axes. Spin-orbit resonances for planet $i$ can be encountered through commensurabilities between $\alpha_i$ and any one of the $g$ frequencies, even if the strength of the perturbation is small in comparison to one of the other modes. This is exemplified by the case of the spin-orbit resonance between Saturn and Neptune\cite{2004AJ....128.2501W, 2004AJ....128.2510H}. 

With the relevant resonant frequencies now defined, we apply these to a set of compact, close-in, nearly coplanar systems to investigate whether planets often exhibit frequency commensurabilities. Our sample consists of 145 planets in 55 multiple-planet systems with masses measured from Transit Timing Variations\cite{2017AJ....154....5H}. We adopted system parameters from Hadden \& Lithwick (2017)\cite{2017AJ....154....5H}; the parameters are provided in their Table 1. We used equations (\ref{alpha}) and (\ref{Darwin-Radau}) along with the assumption of near-synchronous rotation ($\omega \approx n$) to calculate $\alpha$ for these 145 planets, resulting in the dashed histograms in the left panel of Figure 2. The solid line is the distribution of Laplace-Lagrange nodal recession frequencies using equation (\ref{LL g}) for all sets of within-system pairs of planets in the 55 systems, resulting in 141 values in total. 

While the left panel of Figure 2 demonstrates $\lvert g\rvert \sim \alpha$ when the systems are considered in aggregate, it is more useful to compare frequencies for each individual planet in the sample. Here we calculate the range of plausible obliquities each planet could have if captured in spin-orbit resonance. The procedure is as follows. First, for each planet, identify all $g$ frequencies corresponding to secular interactions with the other $n_{pl}-1$ planets in the same system. We assume $e \approx 0$ and an equilibrium rotation rate (for which $d\omega/d t = 0$). In the viscous approach to equilibrium tide theory, this is given by\cite{2007A&A...462L...5L}
\begin{equation}
\frac{\omega_{\mathrm{eq}}}{n} = \frac{2\cos\epsilon}{1+\cos^2\epsilon}.
\label{e=0 equilibrium rotation}
\end{equation}
The only unknown parameters are the planets' Love numbers, which we consider in the range\cite{2011A&A...528A..18K} $k_2 \in [0.1,0.6]$, and the modification to the Laplace-Lagrange frequency due to MMR. We take this in the range $g/g_{\scriptscriptstyle \mathrm{LL}} \in [0.3-1.0]$ for period ratios in $[1.5,1.55]$ or $[2.0,2.05]$. Otherwise, $g/g_{\scriptscriptstyle \mathrm{LL}} = 1$. 

Cassini states are equilibrium solutions of the planet's spin vector that obey the relation,  
\begin{equation}
g\sin(\epsilon-I)+\alpha\cos\epsilon\sin\epsilon=0,
\label{Cassini state relation}
\end{equation}
where $I$ is the inclination of the planet's orbital plane with respect to the invariable plane\cite{2007ApJ...665..754F}. This expression depends on the conditions that the planet's spin angular momentum is much smaller than the orbital angular momentum and that the orbital precession and inclination with respect to the invariable plane are uniform. Correia (2015)\cite{2015A&A...582A..69C} provided a more advanced and generalized theory for defining the Cassini states when these conditions do not hold. The short-period compact systems, however, are still in a regime where the classical approximations suffice for our purposes, so we default to the classical definitions.

Given the small mutual inclinations and large obliquities expected, $I\ll\epsilon$ and equation (\ref{Cassini state relation}) reduces to
\begin{equation}
\lvert{g}\rvert\approx\alpha\cos\epsilon.
\label{resonance condition}
\end{equation}
Using equations (\ref{e=0 equilibrium rotation}) and (\ref{resonance condition}), the obliquity required for spin-orbit resonance is 
\begin{equation}
\label{obliquity at resonance}
\cos\epsilon=\left(\frac{1}{2\alpha_{\mathrm{syn}}/{\lvert{g}\rvert}-1}\right)^{1/2},
\end{equation}
where $\alpha_{\mathrm{syn}}=\alpha(n/\omega)$ is the value of $\alpha$ in the case of synchronous rotation ($\omega=n$). Given the ranges of $k_2$ and $g/g_{\scriptscriptstyle \mathrm{LL}}$, there is a corresponding range in the resonant obliquity. The right panel of Figure 2 depicts this range for each planet. If $\lvert g \rvert > \alpha_{\mathrm{syn}}$ for all values, then no resonance is possible. Among the 145 planets in the sample, $\sim 80\%$ could be in resonance. Among the near-MMR planets, the proportion increases to $\sim 90\%$.

\subsection{1.2 Direct model for orbit and spin evolution.}

In this section and the next, we detail our methodology for modeling the tidal, spin, and orbital evolution of multiple-planet systems. This section outlines our direct numerical integrations using instantaneous accelerations provided by the framework of Mardling \& Lin (2002)\cite{2002ApJ...573..829M}, and the subsequent section discusses secular (orbit-averaged) approximations. We consider a system consisting of a star and $n_{\mathrm{pl}}$ planets, with all bodies endowed with structure. Hierarchical (Jacobi) coordinates are used to calculate the Newtonian orbital evolution. In addition to the standard Newtonian gravitational accelerations, additional accelerations on the bodies due to quadrupolar structure for the gravitating masses, tidal forces, and prescribed disk-migration are also applied. 

The viscoeleastic model for tides posits that the tidal response of a body to gravitational stresses consists of the equilibrium deformation modified by a constant time tracking lag\cite{1907scpa.book.....D, Singer1968, 1979M&P....20..301M, 1981A&A....99..126H}. In this approach, the phase lag angle between the tidal bulge and the line connecting the star's and planet's centers is proportional to the tidal frequency. Mardling \& Lin (2002) adopt equilibrium tide theory as outlined by Eggleton, Kiselva, \& Hut (1998)\cite{1998ApJ...499..853E}, but do not use a constant tidal time lag, $\Delta t$, using instead a constant tidal quality factor, $Q$. We adopt the common constant $\Delta t$ approach. The annual tidal quality factor is then $Q_n = (n\Delta{t})^{-1}$. 

Alternative tidal models offer more complicated relations between the phase lag angle and the tidal forcing frequency. In particular, see the works by Efroimsky \& Williams (2009)\cite{2009CeMDA.104..257E}, Efroimsky (2012)\cite{2012CeMDA.112..283E}, Ferraz-Mello (2013)\cite{2013CeMDA.116..109F}, Correia et al. (2014)\cite{2014A&A...571A..50C}, Storch \& Lai (2014)\cite{2014MNRAS.438.1526S}, Bou\`{e} et al. (2016)\cite{2016CeMDA.126...31B}. Many of these tidal models may be closer to reality than the simple model we have chosen to adopt. Given the absence of super-Earths and sub-Neptunes in the Solar System, however, the rheologies of these planets are highly uncertain. Thus the true dependence of $\Delta t$ on the tidal frequency is unknown. We opted for the most mathematically and physically simple tidal model and also the one that enables the best comparison with past work due to its usage there. Although the fine-grained quantitative details of our results depend on this model choice, the essential traits do not. Bulk dissipation in the planet ensures that any gravitational deformation cannot fully track a source that significantly changes its local sky position on an orbital time scale. A substantial obliquity guarantees this annual motion.

The acceleration on body $i$ due to the quadrupolar gravitational moment of body $j$ is given by
\begin{equation}
\bm{a}_{Q,ji} = \frac{k_{2,j}}{2}\frac{{R_j}^5}{r^4}\left(1+\frac{m_i}{m_j}\right)\left[\left(5(\bm{\omega_j}\cdot\bm{\hat{r}})^2 - \lvert\bm{\omega_j}\rvert^2 - 12\frac{G m_i}{r^3} \right)\bm{\hat{r}} - 2(\bm{\omega_j}\cdot\bm{\hat{r}})\bm{\omega_j}\right].
\end{equation}
Variables that have not yet been defined include $\bm{\omega}_j$, body $j$'s spin vector, and $\bm{r}$, the relative position vector from body $j$ to body $i$. $R_j$ and $m_j$ are, respectively, the radius and mass of body $j$.
We account for accelerations on the planets due to the star's quadrupolar moment, but accelerations due to the planets' quadrupolar moments are negligible. 

The star raises tides in the planets, producing accelerations on them of the form, 
\begin{equation}
\bm{a}_{T,i} =  -\frac{3 n_i k_{2,i} {R_i}^5}{Q_{n,i}}\frac{M_{\star}}{m_i}\frac{{a_i}^3}{r^8}\left[3(\bm{\hat{r}}\cdot\bm{\dot{r}})\bm{\hat{r}} + (\bm{\hat{r}}\times\bm{\dot{r}} - r\bm{\omega_i})\times\bm{\hat{r}}\right].
\end{equation}
We ignore tides raised in the star due to the planets, or tides raised in one planet due to the other. 

Our simulations allow the planets to experience disk-driven orbital migration for some duration (e.g. a few million years) at the beginning of their evolution. We parameterize this using damping accelerations of the form\cite{2013A&A...558A.109A}
\begin{equation}
\bm{a}_{\mathrm{mig},i} = -\frac{\bm{\dot{r}}}{2\tau_{a_i}}. 
\end{equation}
In the absence of resonant interactions between planets, this results in semi-major axis evolution, $\dot{a}_i = a_i/{\tau_{a_i}}$. Parameterized disk-induced eccentricity damping can also be employed\cite{2002ApJ...567..596L, 2013A&A...558A.109A}, but we have not included this mechanism in the model here. We have experimented with doing so, and the results only change in small details.

We follow the evolution of the spin vector of both planets and the star. The spin evolution of the planets 
is given by
\begin{equation}
I_i\bm{\dot{\omega}_i} = -\frac{M_{\star}m_i}{M_{\star} + m_i}\bm{r} \times \left(\bm{a}_{Q,i\star} + \bm{a}_{T,i}\right),
\end{equation}
and that of the star,
\begin{equation}
I_{\star}\bm{\dot{\omega}_{\star}} = \sum_{i=1}^{n_{pl}} -\frac{M_{\star}m_i}{M_{\star} + m_i}\bm{r} \times \bm{a}_{Q,\star i}.
\end{equation}
The quantity $I$ is the fully dimensional moment of inertia.

Provided initial conditions of the system, we evolve the resulting orbital and spin evolution using a Bulirsch-Stoer integrator with the timestep equal to 0.01 times the innermost planet's orbital period and the timestep accuracy parameter set to $\eta=10^{-13}$.

\subsection{1.3 Secular model for orbit and spin evolution.}

The accuracy of the direct integrations outlined above can be checked using secular, orbit-averaged expressions. We adopt the tidal orbital evolution equations from Leconte et al. (2010)\cite{2010A&A...516A..64L}. These are the equilibrium tide expressions of Hut et al. (1981)\cite{1981A&A....99..126H} extended to arbitrary planetary obliquities. We use the secular spin equations of Fabrycky et al. (2007)\cite{2007ApJ...665..754F}, which are in turn adapted from those presented Eggleton \& Kiseleva-Eggleton (2001)\cite{2001ApJ...562.1012E} under the assumption of negligible orbital eccentricities. As in the direct model, we use the constant time lag approximation with the annual tidal quality factor $Q_n = (n\Delta{t})^{-1}$. See also Correia (2009)\cite{2009ApJ...704L...1C} for similar secular expressions to the ones below.

The dissipative secular evolution of each planet's semi-major axis and eccentricity is given by 
\begin{equation}
\gamma_a \equiv \frac{\dot{a}}{a} = \frac{4a}{G M_{\star} m_{\rm p}} K\left[N(e)\cos\epsilon\frac{\omega}{n} - N_a(e)\right] \, ,
\label{gamma_a}
\end{equation}
and
\begin{equation}
\gamma_e \equiv \frac{\dot{e}}{e} = \frac{11a}{G M_{\star} m_{\rm p}}K\left[\Omega_e(e)\cos\epsilon\frac{\omega}{n}-\frac{18}{11}N_e(e)\right]\, ,
\label{gamma_e}
\end{equation}
with $K$ given by 
\begin{equation}
K = \frac{3n}{2}\frac{k_2}{Q_n}\left(\frac{G {M_{\star}}^2}{R_{\rm p}}\right)\left(\frac{R_{\rm p}}{a}\right)^6
\label{tidalK}
\end{equation}
and the following functions of eccentricity:
\begin{align}
N(e) &= \frac{1 + \frac{15}{2}e^2 + \frac{45}{8}e^4 + \frac{5}{16}e^6}{(1-e^2)^6} \\
N_a(e) &= \frac{1 + \frac{31}{2}e^2 + \frac{255}{8}e^4 + \frac{185}{16}e^6 + \frac{25}{64}e^8}{(1-e^2)^{\frac{15}{2}}} \\
\Omega_e(e) &= \frac{1+\frac{3}{2}e^2 + \frac{1}{8}e^4}{(1-e^2)^5} \\
N_e(e) &= \frac{1 + \frac{15}{4}e^2 + \frac{15}{8}e^4 + \frac{5}{64}e^6}{(1-e^2)^{\frac{13}{2}}}.
\end{align}

The secular evolution of each planet's spin-vector may be written as the sum of two torques: a non-dissipative torque due to the star's gravitational force on the oblate figure of the planet, and a dissipative torque due to tides raised in the planet. Explicitly, 
\begin{equation}
\dot{\bm{\omega}} = \dot{\bm{\omega}}_{\mathrm{star}} + \dot{\bm{\omega}}_{\mathrm{tides}}\, ,
\end{equation}
where 
\begin{equation}
\begin{split}
\dot{\bm{\omega}}_{\mathrm{star}} &= \alpha(\bm{\omega}\cdot\bm{\hat{n}})(\bm{\hat{\omega}} \times \bm{\hat{n}}) \\
\dot{\bm{\omega}}_{\mathrm{tides}} &= \frac{M_{\star}m_{\rm p}}{C}\left(\frac{G a}{M_{\star} + m_{\rm p}}\right)^{\frac{1}{2}}\left[-\frac{\bm{\omega}}{2 n t_F} + \frac{1}{t_F}\left(1-\frac{\bm{\omega} \cdot \bm{\hat{n}}}{2n}\right)\bm{\hat{n}}\right].
\label{torques}
\end{split}
\end{equation}
In these expressions, $\bm{\hat{n}}$ is the orbit normal vector, $\alpha$ is the spin-axis precession constant introduced in equation (\ref{alpha}), and $t_F$ is a tidal friction timescale given by
\begin{equation}
t_F = \frac{Q_n}{3k_2}\left(\frac{a}{R_{\rm p}}\right)^5\frac{m_{\rm p}}{M_{\star}}\frac{1}{n}.
\end{equation}

When planets are near MMR, the frequency of the planets' orbital nodal regression deviates from the Laplace-Lagrange secular solution. In these cases, we can still compute the spin dynamics in the secular approximation by using the direct integration orbital evolution within the spin differential equations. That is, the evolutions of $a$ and $\bm{\hat{n}}$ in equations (\ref{torques}) are obtained from the direct integrations. This is the procedure we followed to obtain the secular solutions shown in Figure 3 of the main text.

\subsection{1.4 Construction of the resonant parameter space map.}

The resonant parameter space map in Figure 4 of the main text shows the domain for the inner planet's spin-orbit resonant capture, which is determined by the conditions required for resonant crossing and capture. There are two conditions: (1) $T_{\alpha}/T_g$ must cross unity through above, and (2) the crossing must be adiabatic such that the obliquity is captured in resonance and not just impulsively kicked. 

We consider a set-up in which two planets start slightly wide of exact MMR and migrate inwards convergently into exact resonance. The ratio $T_{\alpha}/T_g$ decreases upon inward migration, so the first condition can be written as two sub-conditions: 
\begin{equation}
(T_{\alpha}/T_g)_{t=0} > 1; \ (T_{\alpha}/T_g)_{t=t_f} < 1.
\end{equation}
Here $t_f$ is the time at which the migration ends. In order to reduce the entire parameter space down to just two key parameters dictating the inner planet's spin-orbit resonance capture ($\tau_{a_1}$ and $a_1(t=t_f)$), we must assign fiducial (and somewhat arbitrary) values for the rest of the parameters. Figure 4 uses the following parameters:
\begin{itemize}
    \item[-] $t_f = 2$ Myr, such that $a_1(t=0) = a_1(t=t_f)\exp({t_f/\tau_{a_1}})$
    \item[-] $(a_2/a_1)_{t=t_f} = 1.5^{2/3}$ (3:2 MMR)
    \item[-] $(a_2/a_1)_{t=0} = 1.05(a_2/a_1)_{t=t_f}$ (5\% wide of 3:2 MMR)
\end{itemize}
Though these parameters must be specified for concreteness, the qualitative features of the figure are the same regardless of their values.

The purple curves in the figure illustrate the $(T_{\alpha}/T_g)_{t=0} > 1$ criterion. Contours of where this equals unity depend on the obliquity. We assume that $\omega = \omega_{\mathrm{eq}}$ and that $\epsilon$ is the obliquity at $t=0$, which could have been pre-excited by an encounter with the stellar oblateness-induced spin-orbit resonance. Explicitly the contours in $\epsilon$ are calculated from
\begin{equation}
\begin{split}
(T_{\alpha}/T_g)_{t=0} &= \frac{\big| g_{\scriptscriptstyle \mathrm{LL}}\big(a_1(t=0), a_2(t=0)\big)\big|}{\alpha_1\big(a_1(t=0), \omega = \omega_{\mathrm{eq}}\big)\cos\epsilon} \\
&= \frac{\big|g_{\scriptscriptstyle \mathrm{LL}}\big(a_1(t=0), a_2(t=0)\big)\big|}{\alpha_1\big(a_1(t=0), \omega = n\big)}\left(\frac{1+\cos^2\epsilon}{2\cos^2\epsilon}\right) = 1.
\end{split}
\end{equation}
Here $g_{\scriptscriptstyle \mathrm{LL}}$ is equation (\ref{LL g}) (which we may assume because the planets are not yet in MMR), and $\alpha_1$ is equation (\ref{alpha}) for planet 1. Note that the dependence of $(T_{\alpha}/T_g)_{t=0}$ on $a_1(t=0)$ and $a_2(t=0)$ is what makes the purple curves depend on the migration timescale, since $a_i(t=0) = a_i(t=t_f)\exp({t_f/\tau_{a_i}})$.

The green curves for the $(T_{\alpha}/T_g)_{t=t_f} < 1$ criterion assume $\epsilon = 0^{\circ}$, as this is the lowest limit. The contours depend on the ratio $g/g_{\scriptscriptstyle \mathrm{LL}}$ since the planets are in MMR. They are calculated from
\begin{equation}
(T_{\alpha}/T_g)_{t=t_f} = \frac{\big|g_{\scriptscriptstyle \mathrm{LL}}\big(a_1(t=t_f), a_2(t=t_f)\big)\big|}{\alpha_1\big(a_1(t=t_f), \omega = n\big)}\left(\frac{g}{g_{\scriptscriptstyle \mathrm{LL}}}\right) = 1.
\end{equation}

Finally, the red dashed curves are the criterion for an adiabatic crossing in the capture direction. This requires that the crossing timescale is slow in comparison to the resonant libration period. Following Hamilton \& Ward (2004)\cite{2004AJ....128.2510H}, the criterion may be expressed as
\begin{equation}
 \dot{\alpha} + \dot{g} \lesssim \alpha \lvert g\rvert\sin\epsilon_0\sin I.
\end{equation}
Here $\epsilon_0$ is the planet's obliquity upon resonant crossing, and $I$ is orbital inclination with respect to the invariable plane, which we calculate using Laplace-Lagrange theory\cite{1999ssd..book.....M}. Also recall that $g<0$. During convergent inward migration, both $\alpha$ and $\lvert g_{\scriptscriptstyle \mathrm{LL}} \rvert$ increase. Given that $\alpha \propto a^{-3}$, the criterion can be reduced to 
\begin{equation}
\frac{3\alpha}{\tau_{a}} + \dot{g} \lesssim \alpha\lvert g \rvert \sin\epsilon_0\sin I.  
\label{adiabatic criterion 1}
\end{equation}
Using that $\lvert g \rvert \approx \alpha\cos\epsilon_0$ at resonant crossing and writing $g$ in terms of $g_{\scriptscriptstyle \mathrm{LL}}$, equation (\ref{adiabatic criterion 1}) becomes
\begin{equation}
\frac{3\lvert g_{\scriptscriptstyle \mathrm{LL}}\rvert}{\tau_{a}\cos\epsilon_0}  +  \dot{g_{\scriptscriptstyle \mathrm{LL}}} +  g_{\scriptscriptstyle \mathrm{LL}}\frac{\frac{d}{dt}(g/g_{\scriptscriptstyle \mathrm{LL}})}{({g/g_{\scriptscriptstyle \mathrm{LL}}})}  \lesssim  {g_{\scriptscriptstyle \mathrm{LL}}}^2\left(\frac{g}{g_{\scriptscriptstyle \mathrm{LL}}}\right)\tan\epsilon_0\sin I.
\label{adiabatic criterion 2}
\end{equation}
This can be simplified a bit further by assuming $\epsilon_0$ and $I$ are small (as this leads to conservative constraints), 
\begin{equation}
\frac{3\lvert g_{\scriptscriptstyle \mathrm{LL}}\rvert}{\tau_{a}}  +  \dot{g_{\scriptscriptstyle \mathrm{LL}}} +  g_{\scriptscriptstyle \mathrm{LL}}\frac{\frac{d}{dt}(g/g_{\scriptscriptstyle \mathrm{LL}})}{({g/g_{\scriptscriptstyle \mathrm{LL}}})}  \lesssim  {g_{\scriptscriptstyle \mathrm{LL}}}^2\left(\frac{g}{g_{\scriptscriptstyle \mathrm{LL}}}\right)\epsilon_0 I.  
\label{adiabatic criterion 3}
\end{equation}
The derivative $\dot{g_{\scriptscriptstyle \mathrm{LL}}}$ is calculated straightforwardly from equation (\ref{LL g}),
\begin{equation}
\dot{g_{\scriptscriptstyle \mathrm{LL}}} = \frac{\mathrm{d}g_{\scriptscriptstyle \mathrm{LL}}}{\mathrm{d}a_1}\dot{a_1} + \frac{\mathrm{d}g_{\scriptscriptstyle \mathrm{LL}}}{\mathrm{d}a_2}\dot{a_2} = 
-\left(\frac{\mathrm{d}g_{\scriptscriptstyle \mathrm{LL}}}{\mathrm{d}a_1}\frac{a_1}{\tau_{a_1}} + \frac{\mathrm{d}g_{\scriptscriptstyle \mathrm{LL}}}{\mathrm{d}a_2}\frac{a_2}{\tau_{a_2}}\right). 
\end{equation}
The derivative of the modification to the Laplace-Lagrange frequency, $\frac{d}{dt}{(g/g_{\scriptscriptstyle \mathrm{LL}})}$, cannot be computed analytically. A rough estimate obtained from the Figure 3 simulation is $\frac{d}{dt}{(g/g_{\scriptscriptstyle \mathrm{LL}})} \approx - 2/\tau_a$, which we use in producing Figure 4. Finally, we use a very conservative value for the obliquity upon resonant crossing, $\epsilon_0 = 5^{\circ}$. Larger values make the adiabatic criterion less stringent. 

~\\

\subsection{\large{Part 2: Background material and broader implications}}

This information is intended to support understanding of the main text and detail the broader implications of our theory. We begin by summarizing the fundamentals of obliquity tides and the dynamics of planets in dissipative Cassini states. Next, we present a simple test integration that demonstrates the efficacy of our direct and secular codes used for modeling planetary orbit and spin evolution. We then upgrade the complexity of the fiducial simulation presented in the main text by accounting for pre-main sequence stellar contraction and spin-down. We show that this stellar evolution can help facilitate the planets' capture in a long-lived, highly oblique state. Finally, we close with a series of consequences and predictions of the theory. We first derive rough constraints on the tidal quality factors of planets in short-period compact systems. We then examine trends in their radii and make a prediction for a dearth of satellite systems.

\subsection{2.1 Cassini states and obliquity tides.}

In the presence of external torques, the plane of a Keplerian orbit of a planet will precess. Relevant torques can stem from a variety of sources, including extended (that is non-point) masses and additional perturbing bodies. The spin dynamics of a planet's rotational pole within the precessing non-inertial frame bear some analogy to the dynamics of a point particle in a rotating potential, such as those found in the circular restricted three-body problem. The points of equilibrium for the spin pole in the rotating frame are known as ``Cassini states''\cite{1966AJ.....71..891C, 2015A&A...582A..69C}. For a dissipationless system, Cassini states require co-planarity of the planetary spin vector, $\bm {\hat{\omega}}$, the planetary orbital momentum unit vector, $\bm {\hat{n}}$ and the total system angular momentum vector, $\bm {\hat{k}}$, and they are the analogs of the Lagrange points of the circular restricted problem. As with the triangular Lagrange points $L_4$ and $L_5$, small perturbations to several of the Cassini states (and in particular, Cassini state 2, which has the direction of the system angular momentum lying between the direction of the spin axis and the orbit normal) generate stable librations.

The viscoeleastic model for tides posits that the tidal response of a body to gravitational stresses consists of the equilibrium deformation modified by a constant tracking lag\cite{1907scpa.book.....D, 1981A&A....99..126H}. In this constant time lag approximation,
for a given eccentricity, $e$, and obliquity, $\epsilon$, the equilibrium rotation rate of the planet (for which $d\omega/d t = 0$) is given by\cite{2007A&A...462L...5L}

\begin{equation}
\frac{\omega_{\mathrm{eq}}}{n} = \frac{N(e)}{\Omega(e)}\frac{2\cos\epsilon}{1+\cos^2\epsilon},
\label{omega eq}
\end{equation}
where $n=2\pi/P$ is the mean-motion, and where $N(e)$ and $\Omega(e)$ are functions of eccentricity,
\begin{align}
N(e) &= \frac{1 + \frac{15}{2}e^2 + \frac{45}{8}e^4 + \frac{5}{16}e^6}{(1-e^2)^6} \\
\Omega(e) &= \frac{1 + 3e^2 + \frac{3}{8}e^4}{(1-e^2)^{\frac{9}{2}}}.
\label{N(e)}
\end{align}

If $e = 0$ and $\epsilon = 0^{\circ}$, then the rotation is synchronous. Supplementary Figure 1 shows the dependence of $\omega_{\mathrm{eq}}/n$ on $e$ and $\epsilon$. At large obliquity, the rotation is sub-synchronous, $\omega_{\mathrm{eq}} < n$, and highly oblique planets have $\omega_{\mathrm{eq}} \ll n$. On an oblique, sub-synchronously rotating planet, an observer at a fixed longitude will see the star drifting ahead in the direction of the planet's rotation (Supplementary Figure 2). The planet's tidal bulge thus lags behind the star, and star-planet torques act to convert orbital energy to heat. To second order in eccentricity, the rate of tidal dissipation in a state of equilibrium rotation is\cite{2007A&A...462L...5L}

\begin{equation}
\frac{\dot{E}_{\mathrm{tide}}(e,\epsilon)}{K} = \frac{2}{1+\cos^2\epsilon}[\sin^2\epsilon + e^2(7+16\sin^2\epsilon)],
\label{dissipation rate}
\end{equation}
where $K$ is given by equation (\ref{tidalK}).

The rate at which orbital energy is converted to heat via tides is a strongly increasing function of obliquity. Supplementary Figure 3 shows the dependence of ${\dot{E}}_{\mathrm{tide}}(e,\epsilon)/{K}$ on $e$ and $\epsilon$. By way of example, for $e=0.01$, the tidal dissipation of energy in the planet is enhanced by a factor of 100 for $\epsilon = 15^{\circ}$ and 1000 for $\epsilon = 45^{\circ}$.

In the presence of steady tidal dissipation within the planet, the locations of the linearly stable Cassini states can be asymmetrically shifted to balance the torques (rather than being linearally destabilized as one might naively expect). This is illustrated with Supplementary Figure 4, which shows a short time evolution of $\bm {\hat{k}}$, $\bm {\hat{\omega}}$, and $\bm {\hat{n}}$, for a planet locked in a dissipative Cassini state 2. The phase shift of $\bm {\hat{\omega}}$ out of the plane defined by $\bm {\hat{k}}$ and $\bm {\hat{n}}$ provides a channel for the planet to steadily dissipate while maintaining a high obliquity. This state of affairs can be brought about if a mechanism operates to sweep the ratio of frequencies, $\vert g\vert/(\alpha\cos\epsilon)$, through unity from above to capture the planet into the Cassini state.

\setcounter{figure}{0}
\renewcommand{\figurename}{Supplementary Figure}

\begin{figure}[H]
\includegraphics[width=0.7\columnwidth]{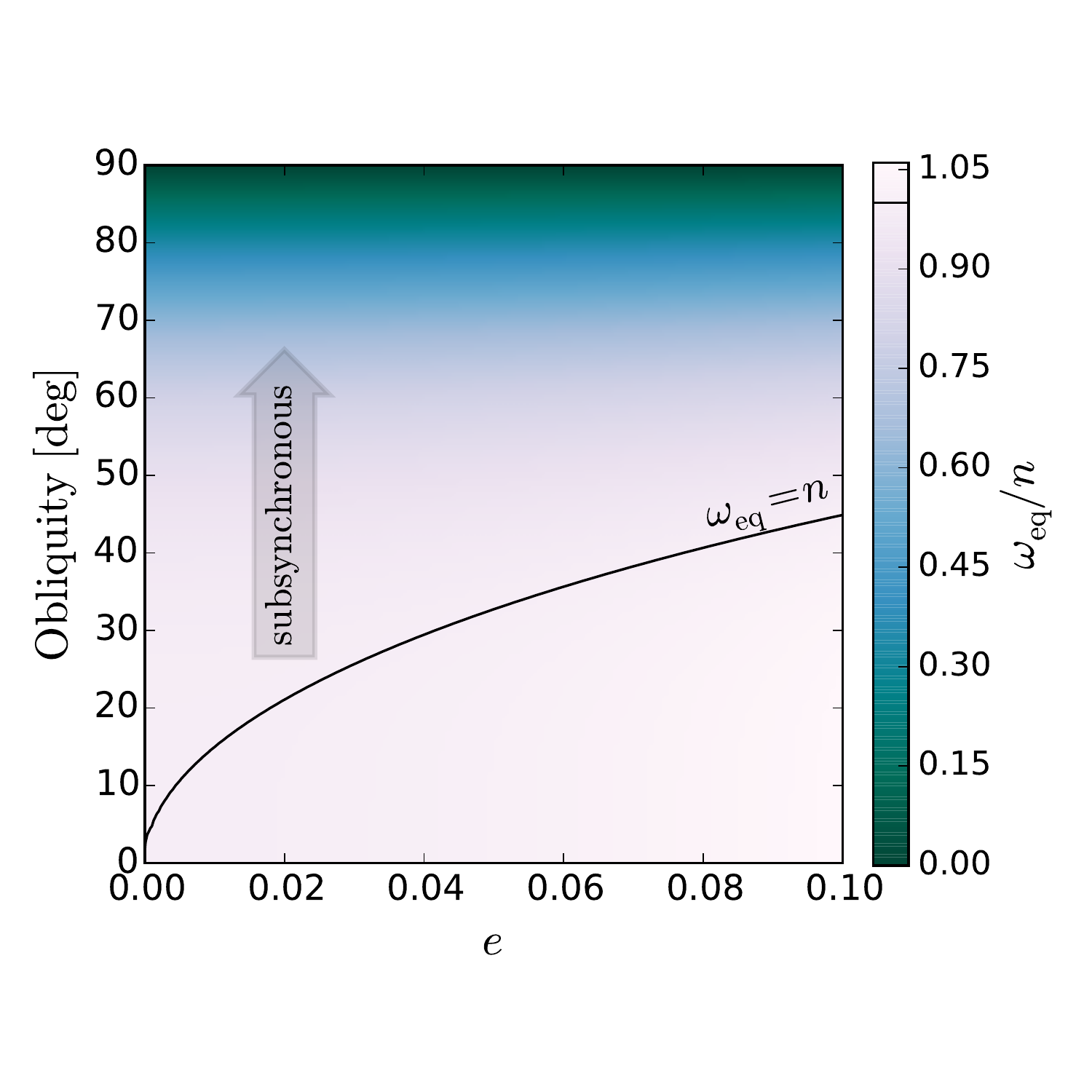}
\caption{\textbf{Equilibrium rotation rate.} \newline
The dependence of the equilibrium rotation rate, $\omega_{\mathrm{eq}}$ (equation (\ref{omega eq})), normalized by the mean-motion, $n=2\pi/P$, as a function of the orbital eccentricity and planetary obliquity. Large obliquities result in sub-synchronous rotation.}
\label{equilibrium rotation rate}
\end{figure}

\begin{figure}[H]
\includegraphics[width=0.55\textwidth]{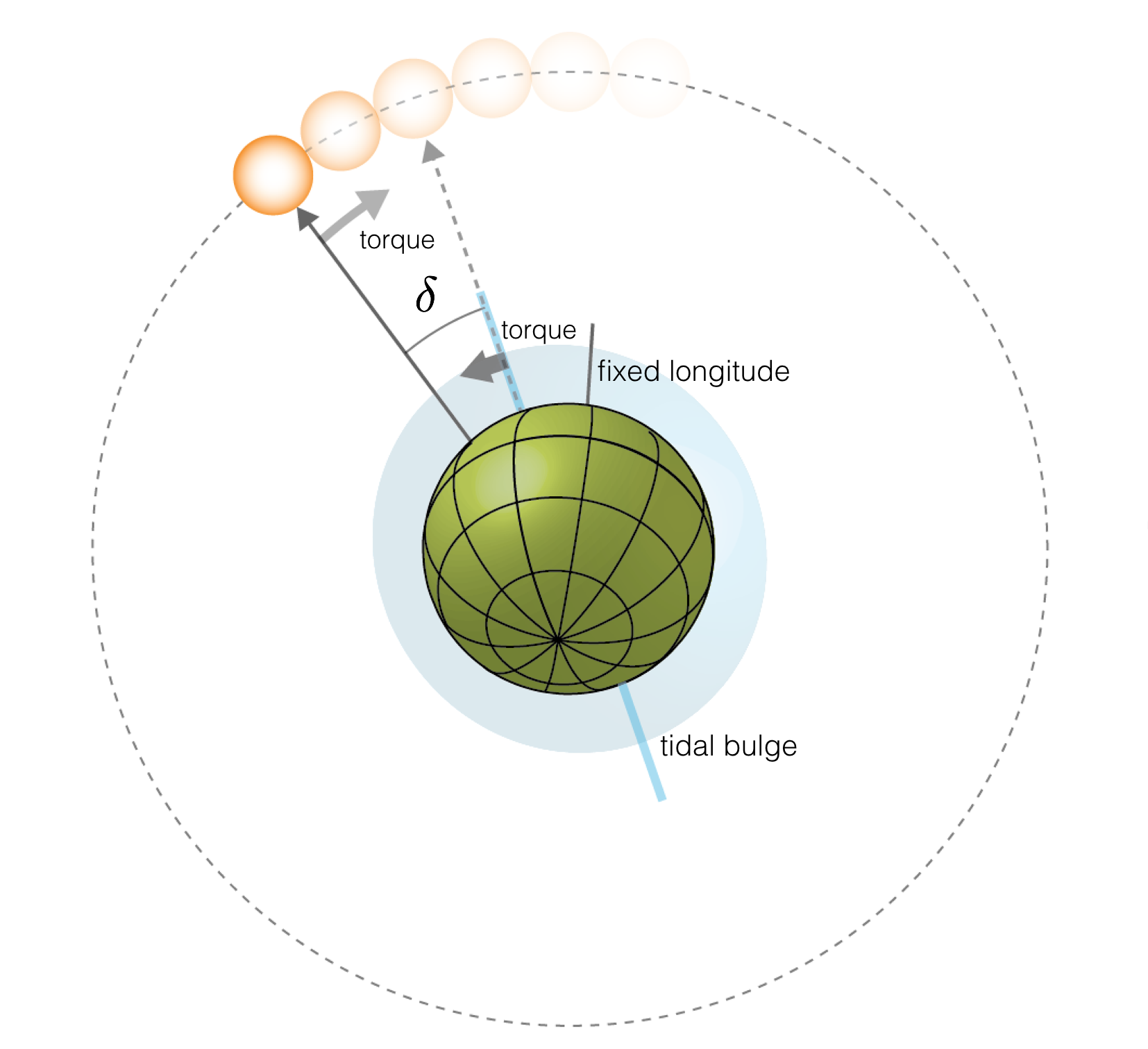}
\caption{\textbf{Tidal dissipation in an oblique sub-synchronous planet.} \newline
When an external mechanism maintains a non-zero obliquity in a dissipative planet, the equilibrium spin rate is $\omega_{\rm eq}/n \sim 2\cos\epsilon/(1+\cos^2{\epsilon})$. The tidal response of the plant has phase lag, $\delta$, with both longitudinal and latitudinal components. Star-planet torques arising from the subsynchronous motion (shown here in a frame rotating at $\omega_{\rm eq}$) act to convert orbital energy into heat.
}
\label{subsynchronous}
\end{figure}

\begin{figure}[H]
\includegraphics[width=0.6\textwidth]{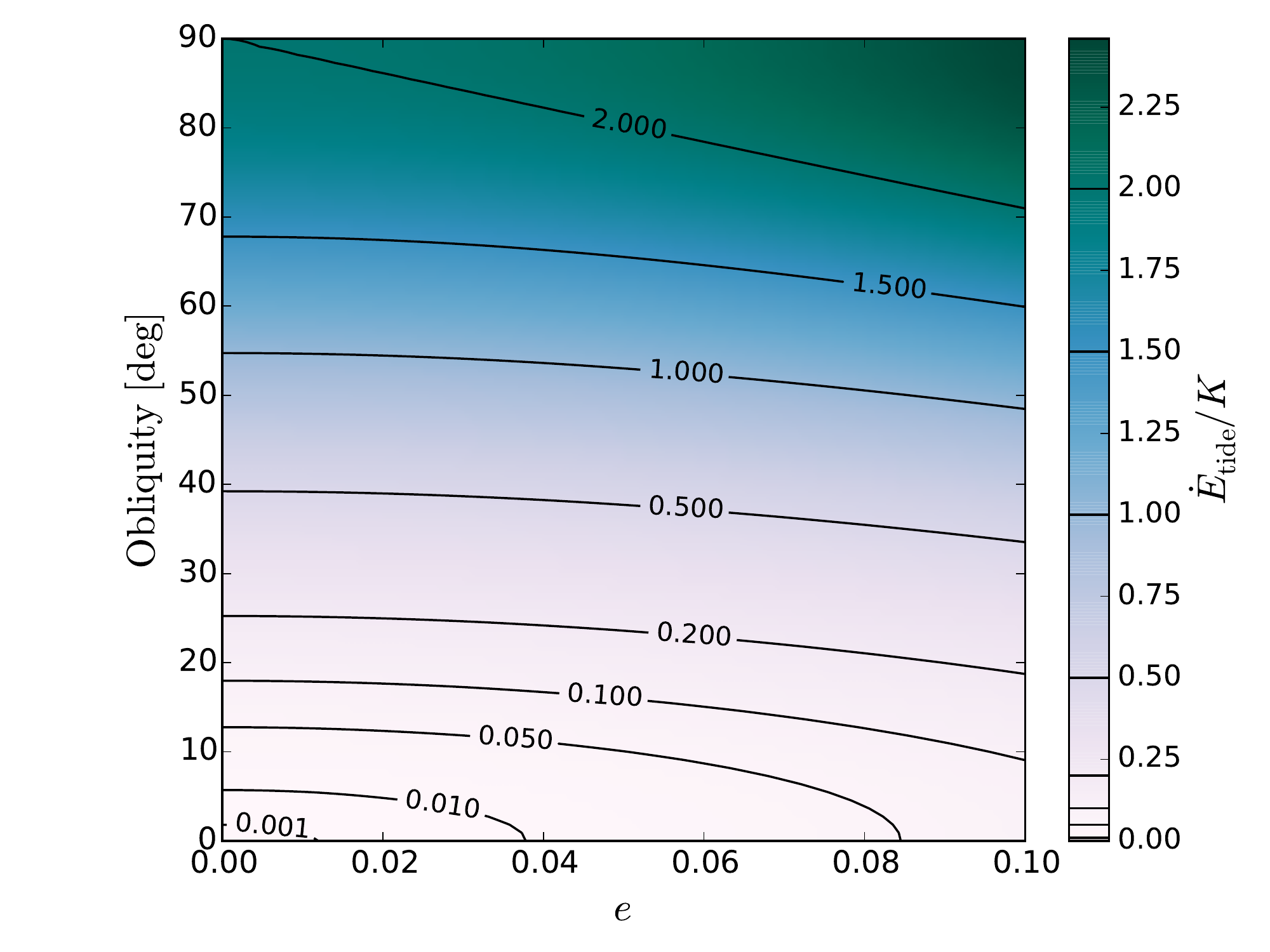}
\caption{\textbf{Tidal dissipation enhancement.} \newline
The normalized tidal dissipation rate, $\dot{E}_{\mathrm{tide}}/K$ (equation (\ref{dissipation rate})), as a function of orbital eccentricity and planetary obliquity. The rate at which energy is dissipated and turned to heat in an oblique planet is enhanced by several orders of magnitude compared to the case with zero obliquity. }
\label{tidal dissipation enhancement}
\end{figure}

\begin{figure}[H]
\includegraphics[width=0.7\textwidth]{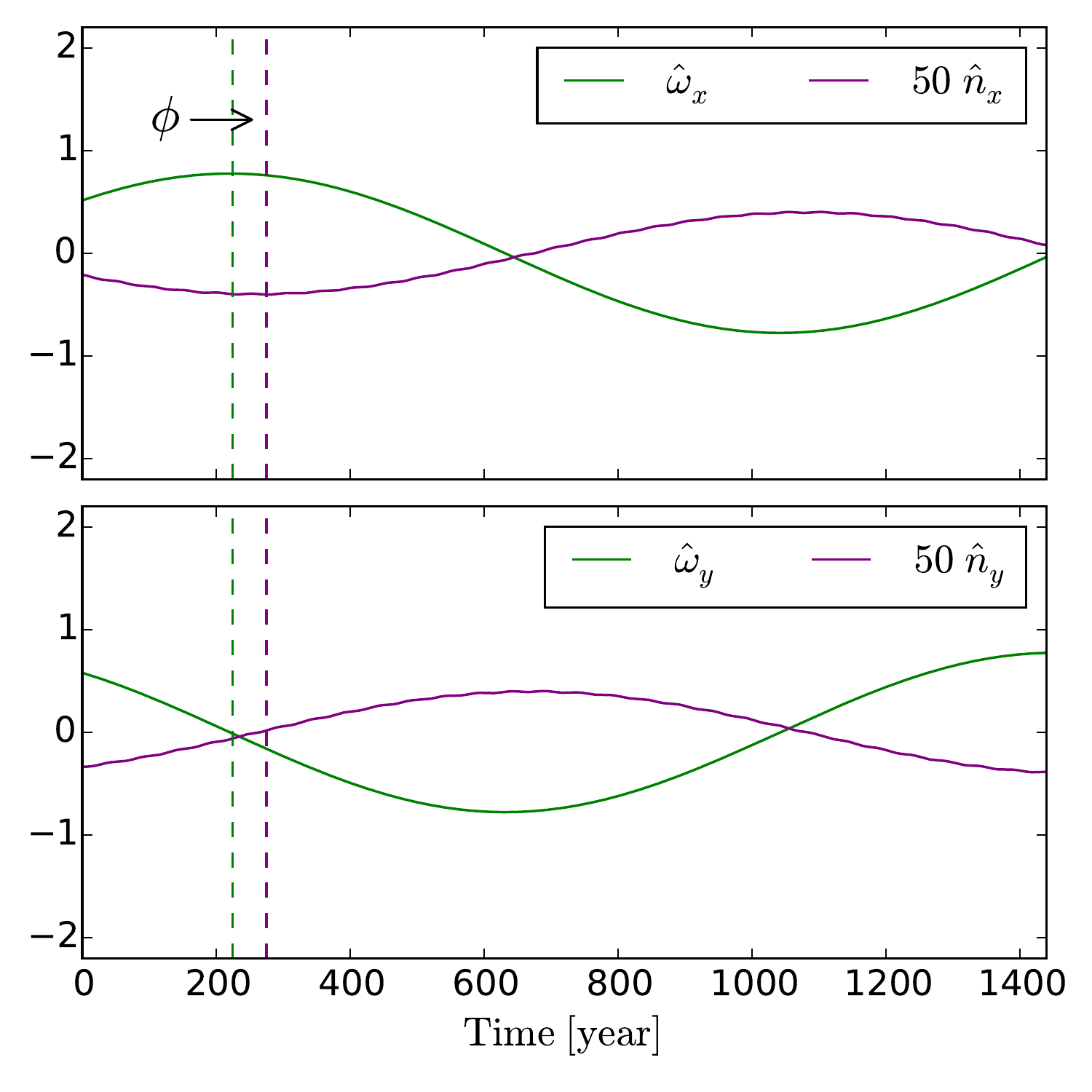}
\caption{\textbf{Cassini state 2 with dissipation.} \newline
Projections of an oblique planet's orbit normal vector, $\bm{\hat{n}}$, and unit spin vector, $\bm{\hat{\omega}}$, onto the invariable plane (the plane perpendicular to the total angular momentum vector, $\bm{\hat{k}}$). $\bm{\hat{n}}$ has been scaled by 50 for better visualization. The top/bottom panels show the x/y components, respectively. This figure was constructed using the simulation displayed in Figure 3. The time segment was taken from the end of the simulation when the two planets were locked in a 3:2 MMR, and the inner planet was trapped in Cassini state 2 with a $52^{\circ}$ obliquity. In a Cassini state 2 with zero dissipation, $\bm{\hat{n}}$, $\bm{\hat{\omega}}$, and $\bm{\hat{k}}$ are coplanar. With dissipation, however, $\bm{\hat{\omega}}$ gets shifted out of the plane. The maximum phase shift before the Cassini state can no longer exist (due to the tidal torque overpowering the perturbation torque)\cite{2007ApJ...665..754F} is $\phi = 90^{\circ}$. Clearly the system in this simulation is far from the limit.}
\label{Cassini state}
\end{figure}

\begin{figure}[H]
\includegraphics[width=\columnwidth]{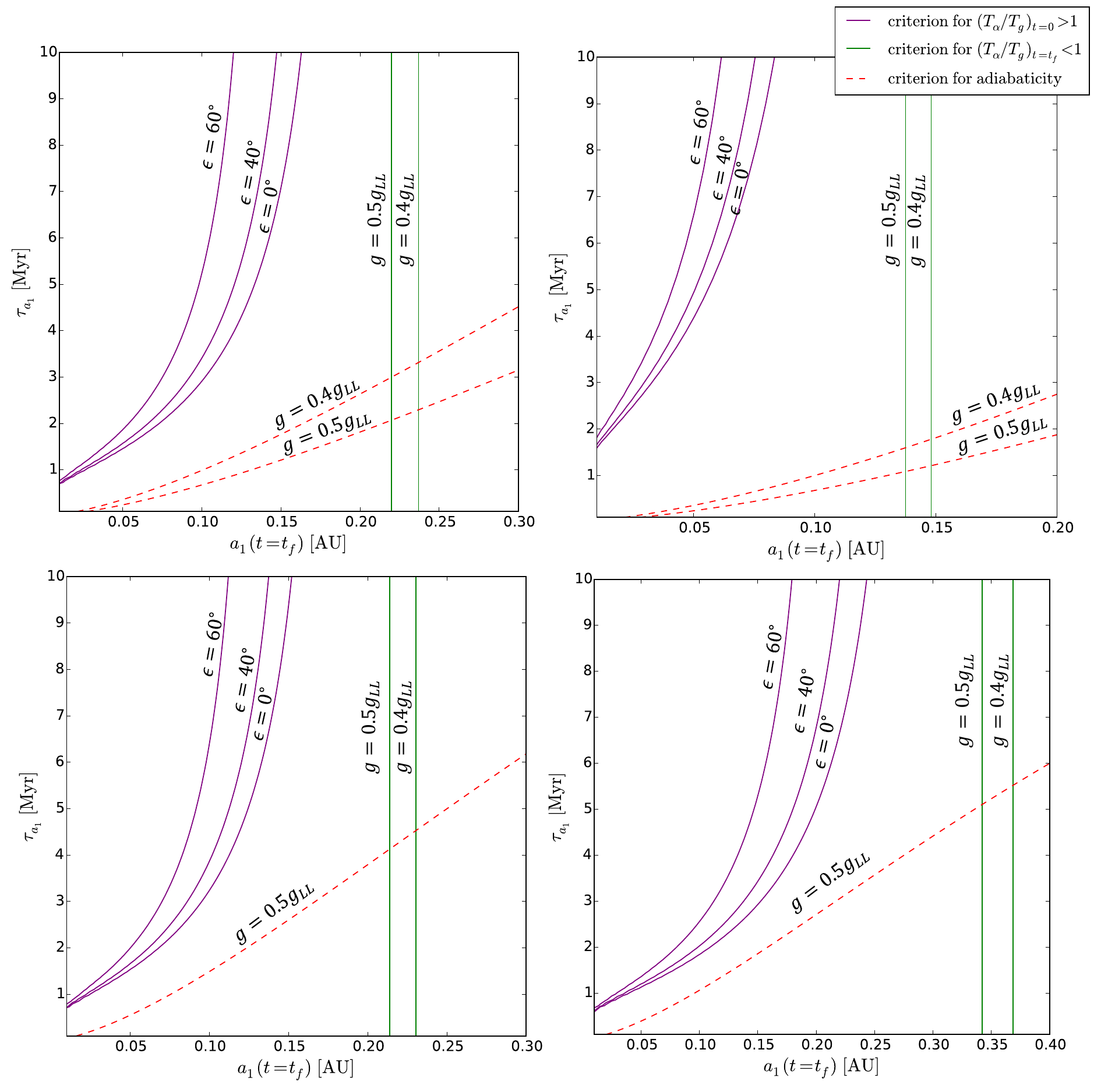}
\caption{\textbf{Domain for the inner planet's spin-orbit resonant capture during convergent inward migration.} \newline
Variations of the parameter space map presented in Figure 4 of the main text. All parameters are identical to those used previously except for the following. \textit{Top left}: $R_1 = R_2 = 4 R_{\oplus}$ rather than $R_1 = R_2 = 2.5 R_{\oplus}$. \textit{Top right}: $t_f = 4$ Myr rather than $t_f = 2$ Myr. \textit{Bottom left}: 2:1 MMR rather than 3:2 MMR. \textit{Bottom right}: 2:1 MMR rather than 3:2 MMR and $R_1 = R_2 = 4 R_{\oplus}$ rather than $R_1 = R_2 = 2.5 R_{\oplus}$. In all cases, the resonant domain is bigger than the case presented in the main text.}
\end{figure}

\newpage 

\subsection{2.2 Spin-orbit resonance in the Solar System.}

We have postulated that spin-orbit resonances play a critical role in shaping compact extrasolar systems. Given this importance, it is worthwhile to review the obliquities of the Solar System planets and satellites and understand why they are/are not involved in spin-orbit resonance. Dynamical clues might result by comparing and contrasting the Solar and extrasolar systems. We credit one of referees of this paper for the idea. 

The obliquity dynamics of many planets in the Solar System are affected by participation in or proximity to spin-orbit resonance. Jupiter\cite{2006ApJ...640L..91W} and Saturn\cite{2004AJ....128.2501W, 2004AJ....128.2510H} are both theorized to be participating in spin-orbit resonances. For Jupiter, the frequency commensurability is between its spin-axis precession and the component of nodal recession due to its interaction with Uranus. For Saturn, the relevant interacting planet is Neptune.  Uranus and Neptune most likely avoided spin-orbit resonances because their spin-axis precession frequencies are too small compared to the orbital frequencies. Similarly, for Earth, the spin-axis precession frequency is too fast (partially because of the Moon's influence) to be involved in any spin-orbit resonances. This is in contrast to the compact extrasolar systems, where $\alpha$ and $\lvert{g}\rvert$ are often close to commensurability.

Former isolated resonances with Neptune (that are no longer active) were possibly influential in the evolution of Venus' obliquity to its present-day, nearly $180^{\circ}$  state\cite{2003Icar..163....1C, 2003Icar..163...24C}. Even more important, however, is the role of chaos, which is thought to have strongly influenced Venus and to still impact Mars today\cite{1993Sci...259.1294T}. Mars' obliquity varies chaotically between $\sim10^{\circ}-50^{\circ}$. These chaotic zones are produced by proximity and overlap of secular spin-orbit resonances. Like the terrestrial planets, chaos may also be active in the spin dynamics of planets in compact extrasolar systems. Our study mainly focused on simple, two-planet systems, but in systems of many planets, large chaotic regions will likely arise as a result of resonance overlap. Chaotic obliquity evolution does not necessarily preclude obliquity tides, however, and may even enhance them due a substantial degree of chaotic wandering of the obliquity. Further research on this topic is required.

Last in our comparison of Solar and extrasolar systems, we make a few remarks regarding the Solar System satellites. The Galilean satellites of Jupiter participate in MMRs and have periods (days to weeks) and satellite-to-planet mass ratios ($M_{\rm s}/M_{\rm J} \sim10^{-4}$) that also suggest $\lvert g \rvert \sim\alpha$. The nodal frequencies of the Galilean satellite orbits are dominated, however, by the recession induced by Jupiter's rotational flattening. With Ganymede, for example, the ratio of $g_{\rm J}$, the frequency of nodal precession arising from Jupiter's oblate figure, to the approximate value, $g_{\rm LL}$, arising from the satellite-satellite secular interactions is approximately
\begin{equation}
\frac{g_{\rm J}}{g_{\rm LL}}  \sim \frac{M_{\rm G}}{M_{\rm J}}\left(\frac{9}{2}J_2\right)^{-1/2}\frac{R_{\rm J}}{a}\sim200.
\label{J2dominates}
\end{equation}
The resulting frequency mismatch prevents capture into secular spin-orbit resonance and precludes the dynamical mechanisms discussed in this letter from playing a role among both the Galilean satellites as well as the other regular satellite families of the Solar System's planets.

\newpage

\subsection{2.3 Simple test integration.}

The direct and secular models may be compared and validated using a test integration that results in tidal evolution of $a$, $e$, $\omega$, and $\epsilon$ on readily observable timescales. Here we consider a simple system consisting of a hot Jupiter orbiting a Solar-like star.  We assigned the planet $m_{\rm p} = M_{\mathrm{Jup}}$, $R_{\rm p} = R_{\mathrm{Jup}}$, $Q = 10^4$, $k_2 = 0.3$, and $C = 0.25$. Its initial period was $P = 3 \ \mathrm{days}$ ($a = 0.04072 \ \mathrm{AU}$) and eccentricity, $e = 0.01$. We gave the planet an initial rotation period of $P_{\rm rot} = 0.5 \ \mathrm{days}$ and obliquity, $\epsilon = 30^{\circ}$. Supplementary Figure 6 shows the resulting orbital and spin-evolution obtained using the direct and secular codes independently. The planet's spin quickly synchronizes and aligns with the orbital normal, and the semi-major axis evolution is most rapid while the obliquity is elevated. Clearly, the agreement between the two integrations is very strong, providing assurance that our direct and secular codes are accurate.  

\begin{figure}[H]
\includegraphics[width=0.8\textwidth]{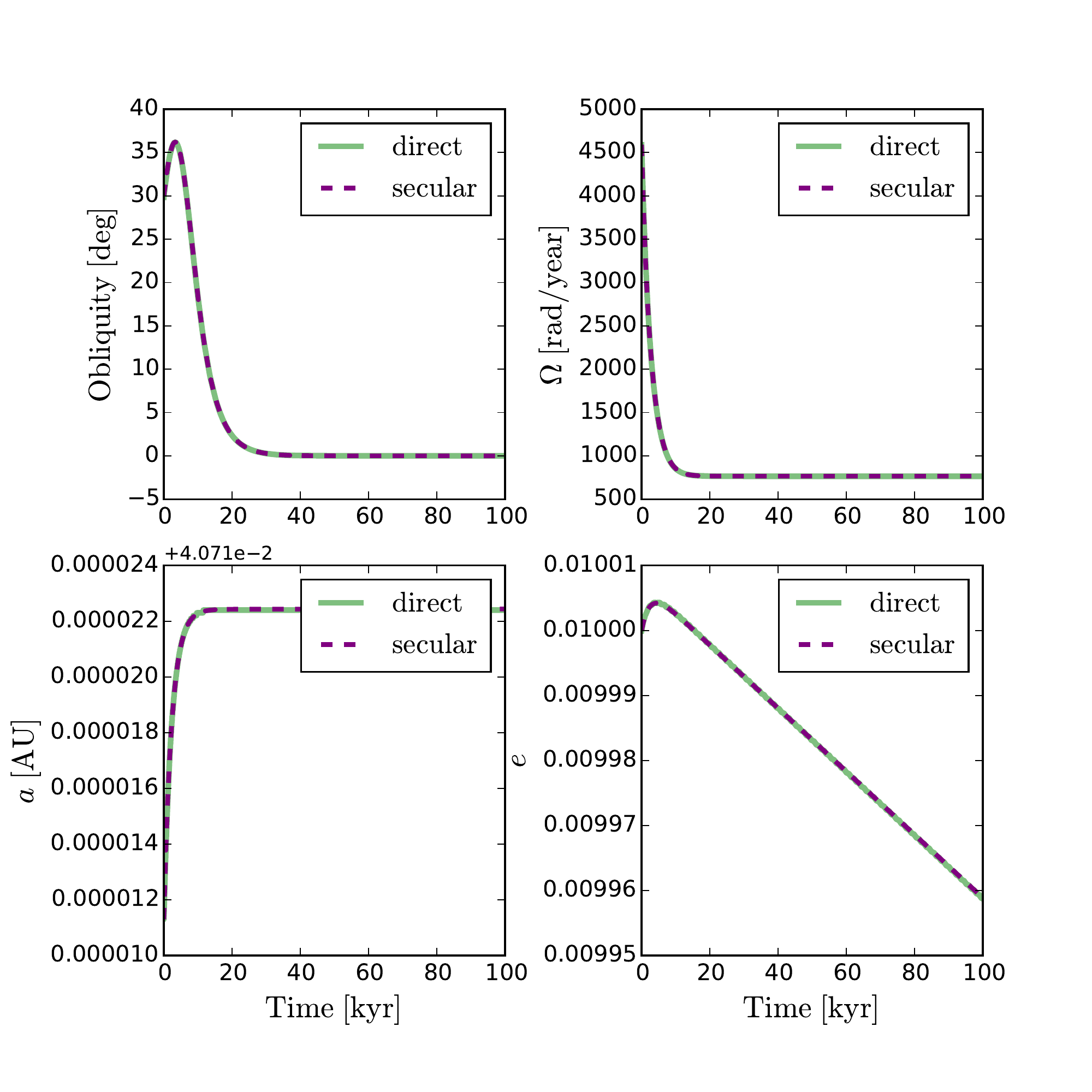}
\caption{\textbf{Test integration of a hot Jupiter.} \newline
The time evolution of the obliquity, spin rate, semi-major axis, and eccentricity of an initially oblique and rapidly-rotating hot Jupiter. We present results computed with both the direct and secular codes.}
\label{HJ simulation results}
\end{figure}

\newpage

\subsection{2.4 Effects of pre-main sequence stellar evolution.}

The fiducial simulation discussed at length in the main text adopted the parameters of a Solar-like main sequence star and did not account for stellar evolution. However, the dynamics we focus on take place early on in the system's lifetime while the star is still young. We must therefore examine how the structural evolution of a young star can affect the planets' obliquity response. Here we show that stellar evolution can not only result in transient resonant locks and obliquity pumping, but can also help facilitate planets to encounter long-term resonant capture. Before concluding the section, we consider the gravitational interactions between the planets and their natal disk, which produces similar effects as those from the young star. 

Pre-main sequence stars have large radii and fast rotation rates\cite{2000stro.book.....T, 2014prpl.conf..433B}, resulting in a significantly enhanced stellar oblateness. The large stellar quadrupole gravitational moment induces a nodal regression of the planetary orbits about the stellar spin vector. In a multiple-planet system, the planets' orbit normal vectors are then perturbed by two components at different frequencies. One frequency, $g_{\star}$, is associated with the large oblateness-induced stellar quadrupole moment and the other, $g_{p-p}$, is due to the planet-planet perturbations. For the type of systems under consideration here, $g_{\star} < g_{p-p}$. Moreover, as the star evolves onto the main sequence, its radius contracts and its rotation rate slows, such that $g_{\star}$ gets progressively smaller.   

A planet's spin axis can be trapped into a Cassini state with either $g_{\star}$ or $g_{p-p}$, depending on which resonance is encountered first and whether the conditions are right for capture. Planets in typical short-period compact systems are readily captured into resonances with $g_{\star}$. Due to the stellar contraction and spin down that reduces $g_{\star}$ to small values, a resonant lock with this frequency can result in the planet's obliquity being forced to nearly $90^{\circ}$. Eventually, $g_{\star}$ becomes so small that the transient resonant lock is inevitably broken and the excited obliquity damps back down. The damping causes $T_{\alpha}$ to decrease rapidly. If $T_{\alpha}/T_{g_{p-p}}$ approaches unity from above, the planet can be locked into a new Cassini state with $g_{p-p}$ and captured into a long-term, large obliquity state. In other words, the transient obliquity excitation produced by the $g_{\star}$ resonance tends to ensure that the $g_{p-p}$ resonance is approached in the correct direction for capture. This is important because in some cases, the $g_{p-p}$ resonance would either never be approached or be encountered in the wrong direction save for the prior encounter with the $g_{\star}$ resonance. 

We illustrate these dynamics with an example simulation presented in Supplementary Figure 7. The structural parameters of the planets, migration timescale and duration, initial eccentricities, initial obliquities, and rotation periods were identical to those of the Figure 3 fiducial simulation. The initial semi-major axes of the planets were slightly smaller, $a_1(0) = 0.0923 \ \mathrm{AU}$ ($P_1(0) = 10.2 \ \mathrm{days}$) and $a_2(0) = 0.1242 \ \mathrm{AU}$ ($P_2(0) = 16.0 \ \mathrm{days}$). These smaller values were selected to emphasize the ubiquity of the capture mechanism at various orbital separations. The evolution of the stellar radius was parameterized using
\begin{equation}
R_{\star}(t) = R_{\star}(0)\left(1 + \frac{t}{\tau_{R_{\star}}}\right)^{-1/3},
\end{equation}
which is derived by assuming Kelvin-Helmholtz contraction of a polytropic body\cite{2013ApJ...778..169B, 2017AJ....154...93S}. 
We took $R_{\star}(0) = 3 R_{\odot}$ and $\tau_{R_{\star}} = 1 \ \mathrm{Myr}$. The stellar spin down was parameterized using 
\begin{equation}
\omega_{\star}(t) = \omega_{\star}(0) + \left[\omega_{{\star},\min} - \omega_{\star}(0)\right]\left[1-\exp\left(-\frac{t}{\tau_{\omega_{\star}}}\right)\right].
\end{equation}
We adopted $\omega_{\star}(0) = 2\pi \ \mathrm{days}^{-1}$, $\omega_{\star,\min} = \pi/10 \ \mathrm{days}^{-1}$, and $\tau_{\omega_{\star}} = 1 \ \mathrm{Myr}$. The important features of the system's evolution are not sensitive to the specific functional forms or parameter values adopted for $R_{\star}(t)$ and $\omega_{\star}(t)$. As with the simulation presented in the main text, we emphasize that specific parameter values must be chosen for concreteness, but the resulting dynamics are robust and common to a wide range of parameters.

The results of the simulation are shown in Supplementary Figure 7. Both planets are trapped into a Cassini state with $g_{\star}$ and have their obliquities rapidly excited. The rate of increase of $T_{g_{\star}}$ eventually exceeds adiabaticity limits and breaks the resonant locks, but as the obliquities damp back down, $T_{\alpha,1}$ and $T_{\alpha,2}$ approach $T_{g_{p-p}}$ from above. Though the inner planet misses the resonance due to non-adiabaticity, the outer planet is trapped in a stable long-lived resonance with $\epsilon \sim 65^{\circ}$. Note that the planets would not have encountered the $g_{p-p}$ resonance without the initial forcing from the stellar oblateness.

In addition to interactions with the stellar gravitational quadrupole, young planetary systems will also be subject to secular planet-disk interactions. The quadrupole potential of the disk induces yet another component of nodal recession for the planets\cite{2017AJ....154...93S}. Like the stellar oblateness-induced recession, the disk-induced nodal recession rate decreases in magnitude as the system evolves, here corresponding to the gas disk dissipating away. The forcing on the planets from the disk therefore behaves similarly to that from the young oblate star. It provides an additional opportunity for transient resonant locks that ultimately help set the planets up for long-term capture in the $g_{p-p}$ resonance.

\begin{figure}
\includegraphics[width=0.73\textwidth]{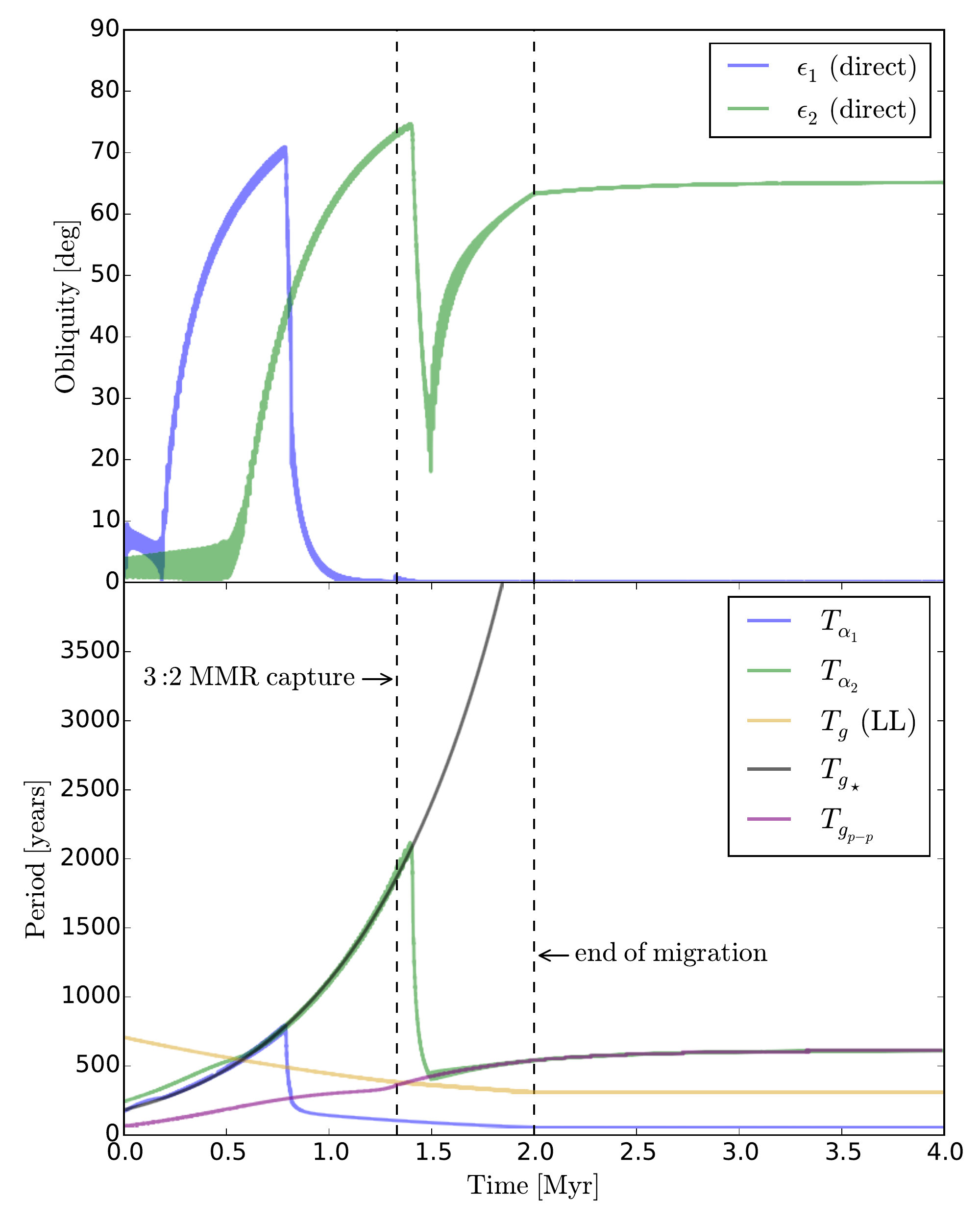}
\caption{\textbf{Simulation including pre-main sequence stellar evolution.} \newline
Dynamical evolution of a fiducial, two-planet system in which the initially large and rapidly rotating young star undergoes contraction and spin-down and the planets experience (imposed) inward, convergent orbital migration for the first $2 \ \mathrm{Myr}$. The 3:2 MMR is captured at $\sim 1.3 \ \mathrm{Myr}$. \textit{Top panel}: Evolution of the obliquities with the direct code. \textit{Bottom panel}: The blue and green curves represent the precession periods of the inner and outer planets, respectively. The yellow curve indicates the uniform nodal recession period according to Laplace-Lagrange theory. A numerical FFT decomposition of the planets' nodal time series computed the two frequencies, $g_{\star}$ and $g_{p-p}$, composing the two components of the planets' orbit normal vector evolution. (The frequencies are the same for each planet.) 
}
\label{stellar evolution sim}
\end{figure}

\newpage

\subsection{2.5 Constraints on the tidal quality factor.}
In the scenario under discussion, many planet pairs with period ratios just wide of a first-order MMR have evolved to their current locations through a combination of resonant forcing and dissipation that damps eccentricities. The distance traveled beyond resonance is related to the amount of dissipation experienced by the planets over the system age, which in turn is directly connected to their tidal quality factors, $Q$. ($Q$ is defined as the inverse of the fractional energy dissipated per tidal oscillation cycle.) Here we use this connection to obtain population-level, order-of-magnitude constraints on the $Q$ values of planets in compact multiple-planet systems.

Once a long-term dissipative equilibrium has been established, the annual tidal quality factor is related (to second order in eccentricity) to the planet's physical and orbital properties through
\begin{equation}
Q_n={6k_2 a n}\,{\left(\frac{da}{dt}\right)}^{-1} \left(\frac{M_{\star}}{m_{\rm p}}\right)\left(\frac{R_{\rm p}}{a}\right)^{5}\left[\frac{\sin^2 \epsilon + e^2(7+16\sin^2 \epsilon)}{1+\cos^{2} \epsilon}\right]\, .
\end{equation}
The nodal frequency, $g$, undergoes significant variations during the formation and evolutionary phases of a planetary system. As a consequence, it is expected that for systems that are captured into secular spin-orbit resonance, $g$ will undergo substantial post-capture evolution. The constraint that $T_{\alpha}/T_g$ remain at unity thus implies that significant obliquities $\epsilon \gtrapprox 30^{\circ}$ will be commonplace.

For near-circular orbits, the obliquity-boost factor thus reduces to $\sin^2\epsilon/(1+\cos^2 \epsilon)\sim 1/2$ for a wide range of axial tilts. The expression thus simplifies to
\begin{equation}
Q_n \sim n \tau_{\rm{age}}\left(\frac{a}{\delta a}\right)\frac{M_{\star}}{m_{\rm p}} \left(\frac{R_{\rm p}}{a}\right)^5\, ,
\label{approxQ}
\end{equation}
where we adopt $k_2=1/3$, and identify $da/dt\sim\delta a/\tau_{\rm {age}}$, where $\delta a$ is the distance migrated during the resonant repulsion process, and $\tau_{\rm {age}}$ is the age of the system. For a fiducial dissipating planet with $M_{\star}=M_{\odot}$, $m_{\rm p}=5M_{\oplus}$, $P=10\,{\rm {days}}$, and $R_{\rm p}=2.5R_{\oplus}$, and $\tau_{\rm {age}}=5\,{\rm Gyr}$, we find $Q\sim5\times10^{4}$, which is similar to the $Q$ values inferred for Uranus\cite{1990Icar...85..394T} and Neptune\cite{2008Icar..193..267Z} in our own Solar System and is consistent with the few published extrasolar super-Earths/sub-Neptunes $Q$ estimates\cite{2017AJ....153...86M, 2018AJ....155..157P}.

Supplementary Figure 8 shows the distribution of $Q_n$ values estimated using equation (\ref{approxQ}) for transiting planets in the NASA Exoplanet Archive (https://exoplanetarchive.ipac.caltech.edu/) that are members of adjacent pairs having $1.5<P_1/P_2<1.53$ and $2<P_1/P_2<2.07$. Tabulated planetary masses, stellar masses, and stellar ages are adopted when available, otherwise we use default values $m_{\rm p}=5\,M_{\oplus}$, $M_{\star}=M_{\odot}$, and $\tau_{\rm {age}}=5\,{\rm Gyr}$. For pairs with period ratios in the ranges of interest, it is not known which planet(s) (if either) are responsible for obliquity-driven repulsion. Moreover, we expect that some of the systems with period ratios emblematic of repulsion lie only coincidentally near the commensurability. We thus adopt
\begin{equation}
\frac{\delta a}{a} \sim 1-\left[\frac{(k+1)}{k}\frac{P_1}{P_2}\right]^{2/3}
\end{equation}
for every planet, with the understanding that this relation will be applicable for roughly 1/3 to 1/2 of all planets considered. 

We expect that the planets that are \textit{not} responsible for the dissipation will exhibit an approximately uniform distribution of spurious and generally low $Q$ values.
Simultaneously, there should be a pronounced peak of true $Q$ estimates associated with planets that are experiencing (or have experienced) significant long-term obliquity dissipation. The location of the peak at $Q\sim10^4$ within the data thus constitutes a direct measurement of the  typical dissipation efficiency of members of the \textit{Kepler}-multiple planet population, and is, by extension, an important clue to their bulk geophysical properties. For comparison with the giant planets of the Solar System, Supplementary Figure 8 marks the current $Q$ estimates for Earth\cite{1999ssd..book.....M} and Jupiter\cite{1977Icar...32..443G, 2009Natur.459..957L}, as well as lower limits for Uranus\cite{1990Icar...85..394T} and Neptune\cite{2008Icar..193..267Z}. 

\begin{figure}[H]
\includegraphics[width=0.73\textwidth]{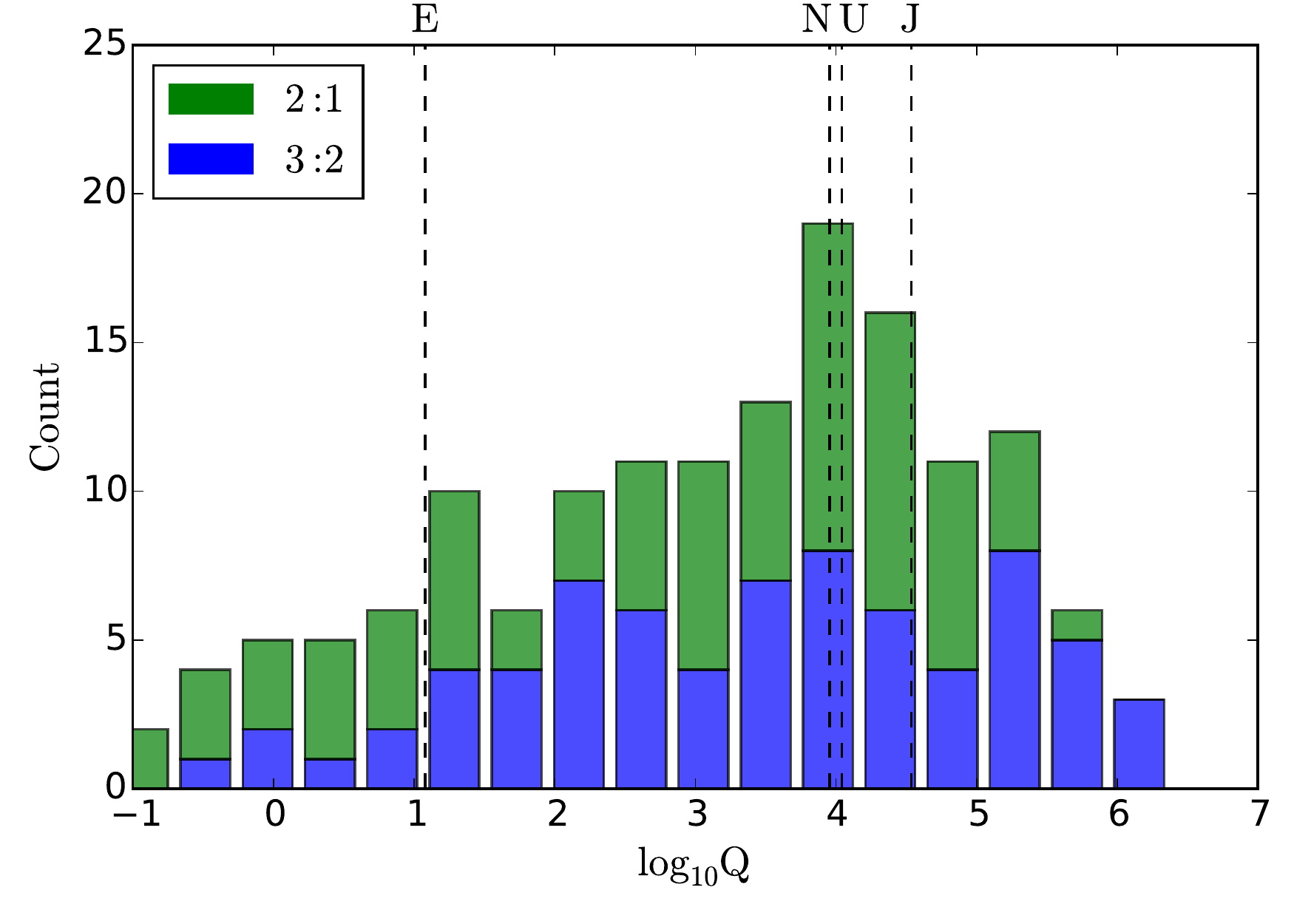}
\caption{\textbf{Estimates of the tidal dissipation quality factor, $Q$, of super-Earths/sub-Neptunes.} \newline
The sample is transiting planets near the 3:2 and 2:1 commensurabilities in the NASA Exoplanet Archive, for which measurements of the stellar age, the planetary mass, and the planetary radius, are all available. The $Q$ values peak at $10^{4}$, a value that is similar to Uranus and Neptune, but lower than Jupiter and higher than the terrestrial planets. The vertical dashed lines mark the $Q$ estimates for Earth and Jupiter, as well as lower limits for Uranus and Neptune.}
\label{Qhist}
\end{figure}

\newpage

\subsection{2.6 Testable predictions: planetary radii.}

Independent of the obliquity and eccentricity, the efficiency of tidal heating depends strongly on the planetary radius, with the dissipation rate of equation (\ref{dissipation rate}) containing an $\dot{E}_{\mathrm{tide}}\propto {R_{\rm {p}}}^5/a^6$ dependence. This has several consequences if obliquity tides are responsible for resonant repulsion. As discussed in the main text, planets that exhibit resonant repulsion should (statistically) be larger than those which have not. As shown in Supplementary Figure 9, when tested with transiting planets from the NASA Exoplanet Archive that are members of adjacent pairs, this is clearly the case. To investigate the statistical significance of the difference, we used Mood's median test. We split the sample into two sub-samples. The first sub-sample comprised all planets in pairs for which $1.405 < P_2/P_1 < 1.5$ or $1.9 < P_2/P_1 < 2.0$, and the second, all planets in pairs for which $1.5 < P_2/P_1 < 1.53$ or $2< P_2/P_1 < 2.07$. The median planet radius of the first sub-sample is 0.14$R_{\mathrm{Jup}}$ and the second 0.21$R_{\mathrm{Jup}}$. An application of Mood's median test to these sub-samples finds the difference in the medians to be significant with a p-value, $p < 6\times10^{-5}$.

Similar to the expectation that repulsed pairs should harbor planets with generally larger radii, the strong radial dependence of $\dot{E}_{\mathrm{tide}}$ on $a$ should generate a $R_{\rm p}(a)$ relation for repulsed planets that is steeper than the $R_{\rm p}(a)$ relation for the ``control'' group consisting of members of pairs that lie interior to the $(k+1):k$ commensurabilities. Supplementary Figure 10 indicates that this relation is present for both the 2:1 and 3:2 populations. We split the sample of planets into four sub-samples and calculated least squares linear fits to $R_{\rm p}/R_{\mathrm{Jup}}$ vs. $\log_{10}(P/\mathrm{days})$. For planets in pairs with $1.41<P_2/P_1<1.5$, the slope is $0.113 \pm 0.027$, whereas for planets in pairs with $1.5 < P_2/P_1 < 1.53$, it is $0.169 \pm 0.045$. As for planets near the 2:1 commensurability, for planets in pairs with $1.9 < P_2/P_1 < 2.0$, the slope is $0.188 \pm 0.037$. Finally, for planets with $2 < P_2/P_1 < 2.07$, it is $0.254 \pm 0.050$. For both the near 2:1 and near 3:2 cases, the difference in slopes for pairs short of resonance and pairs wide of resonance is $\sim1-2\sigma$. While this is not definitive in and of itself, it is interesting to note that the trend matches expectations. If the effect is indeed real, its statistical significance will be improved as additional near-resonant pairs of transiting planets are detected with the forthcoming TESS mission.

\begin{figure}
\includegraphics[width=0.73\textwidth]{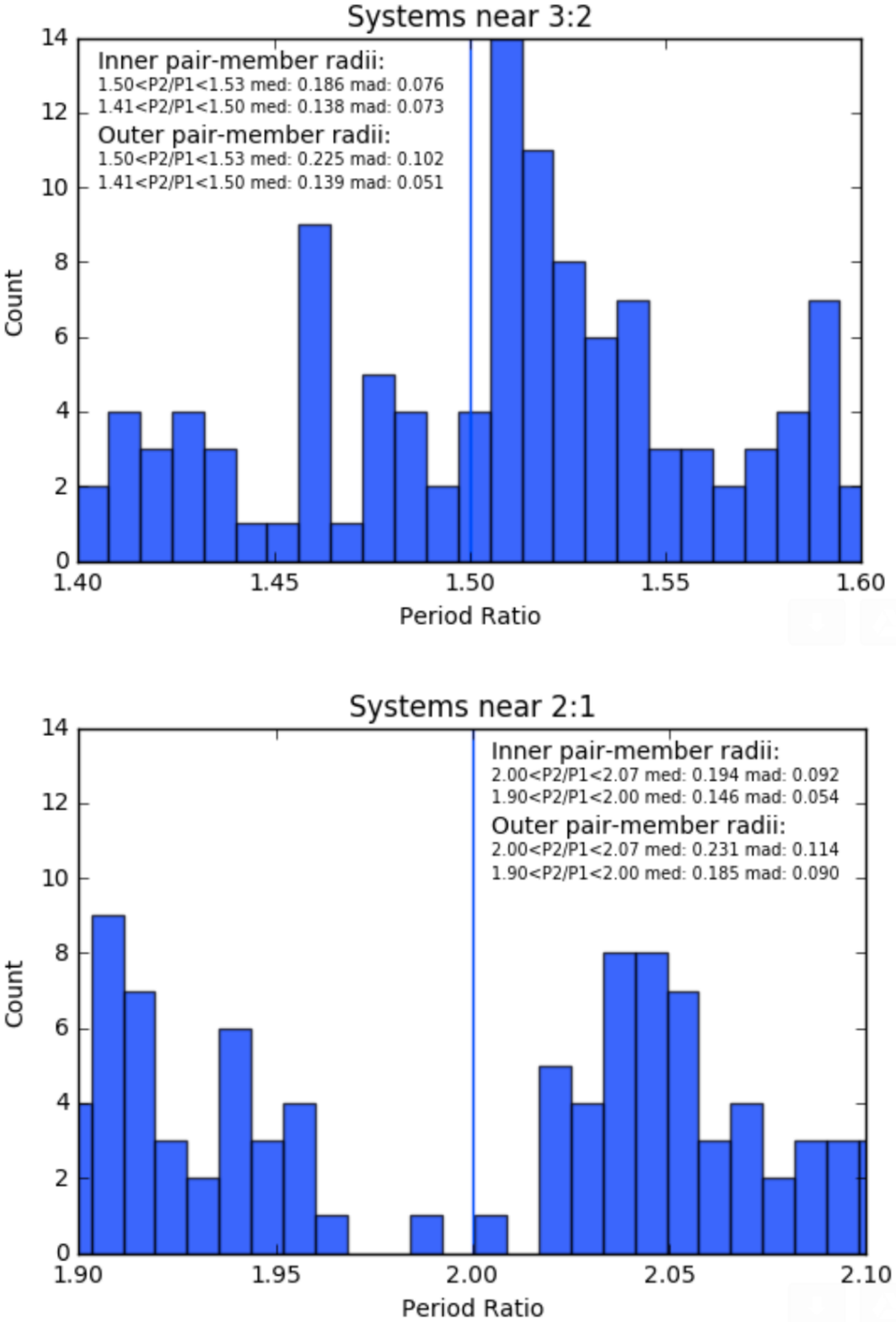}
\caption{\textbf{Period ratio distributions.} \newline
Period ratios for adjacent confirmed planets with measured radii in multiple-planet systems drawn from the NASA Exoplanet Archive. }
\label{Period Ratio Histograms}
\end{figure}

\begin{figure}
\includegraphics[width=0.73\textwidth]{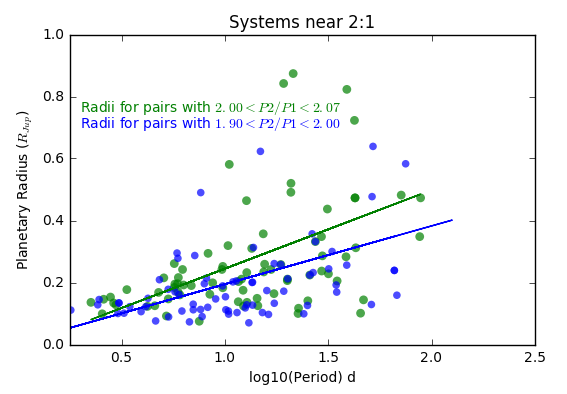}
\label{temporaryFitFig1}
\end{figure}

\begin{figure}
\includegraphics[width=0.73\textwidth]{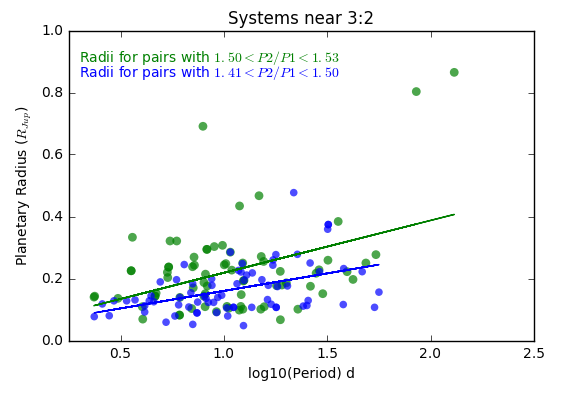}
\caption{\textbf{Radius--period relation for planets near MMR.} \newline
For planets close to the 2:1 (\textit{top panel}) and 3:2 (\textit{bottom panel}) commensurabilities, $R_{\rm {P}}(P)$ increases more steeply for planets belonging to pairs that are wide of the resonance (\textit{green points}) in comparison to planets belonging to pairs that are short of the resonance (\textit{blue points}). As discussed in the text, we expect that only 1/3 to 1/2 of the planets marked with green points are undergoing obliquity-driven dissipation. }
\label{temporaryFitFig2}
\end{figure}

\newpage

\subsection{2.7 Testable Predictions: Paucity of satellites.}

We close this section by discussing predictions of our theory with regard to satellites of planets in short-period compact systems. Planetary satellites, if they are present, provide a significant alteration of the spin dynamics of the planets, and thus provide an opportunity for a consistency test of our hypothesis that obliquity tides are responsible for driving planetary pairs to period ratios, $P_2/P_1>(k+1)/k$. When a satellite's nodal precession frequency in a planet's equatorial plane is rapid in comparison to the planet's spin frequency, the satellite maintains constant inclination relative to the planetary equator, and the satellite-planet system functions as a unit\cite{1965AJ.....70....5G}. The planetary precession rate is modified to be\cite{2004AJ....128.2501W} 
\begin{equation}
{\alpha}=\frac{3n^2(J_2+q)}{2\omega(C+l)}\, .
\end{equation}
The quantity $q$ is the effective quadrupole coefficient of the satellites 
\begin{equation}
    q=\frac{1}{2}\sum_{j}\frac{ \frac{m_j}{m_{\rm p}}\left(\frac{a_j}{R_{\rm p}}\right)^{2}\sin(\epsilon-i)}{\sin \epsilon}\,,
\end{equation}
where $j$ indexes the masses and semi-major axes of the satellites, and $i$ is the inclination of the planet's orbit relative to the planetary system's orbital plane, and 
\begin{equation}
l=\sum_{j} \frac{m_j}{m_{\rm p}}\left(\frac{a_j}{R_{\rm p}}\right)^{2}\frac{n_j}{\omega}\, ,
\end{equation}
is the ratio of angular momentum in the satellite system to the normalizing angular momentum factor $m_{\rm p}{R_{\rm p}}^2\,\omega$.

Within the \textit{Kepler}-detected sample of multiple-planet systems, a typical sub-Neptune class planet belonging to a near-resonant pair with $P_1/P_2>(k+1)/k$  has radius $R_{\rm p}\simeq2.5\,R_{\oplus}$ and period $P\simeq10\,{\rm {days}}$. Adopting a mass, $m_{\rm p}=5\,M_{\oplus}$, and Love number $k_2=0.3$, this gives $J_2=2\times10^{-5}$. The regular satellite systems of the Jovian planets in our own Solar System display a scaling property such that $m_{\rm sat}/m_{\rm p}\sim10^{-4}$, and  $P_{\rm sat}\sim10\,{\rm {days}}$. It is possible that the planet formation process that gives rise to super-Earths/sub-Neptunes also yields satellite systems of this type, although because of the smaller Hill radii, $R_{H}=a_{\rm p}(m_{\rm p}/3M_{\star})^{1/3}$, they would, in some cases, need to be more compact than the implied scaling in order to be dynamically stable. A putative satellite of the regular Jovian type orbiting a typical super-Earth/sub-Neptune would have $q\sim5\times10^{-3}$, implying a ratio $q/J_2>100$. By contrast, $l \ll C$ for such a satellite system, implying that the presence of significant moons would increase the spin precession rate, $\alpha$, by several orders of magnitude. The resulting large-scale mismatch with the characteristic nodal frequencies, $g$, would render capture into secular spin-orbit resonance far less likely. As a consequence, we expect that the population of resonant-repulsed \textit{Kepler} pairs will not generally have satellites. Although individual exomoon candidates remain unconfirmed, our prediction can be statistically tested through the stacking of phase-folded transits of the near-resonant populations\cite{2018AJ....155...36T}. This exercise could be carried out with the current census of \textit{Kepler}-detected planets, and potentially extended to high signal-to-noise by augmenting with data obtained by TESS.

\section*{\large{Data availability}}
The data that support the plots within this paper and other findings of this study are available from the corresponding author upon reasonable request.

\end{document}